\documentclass[pra, twocolumn, preprintnumbers, amsmath, amssymb, nofootinbib, floatfix]{revtex4}

\usepackage{graphicx, bm, tikz}
\usepackage[breaklinks]{hyperref}

\makeatletter
\def\graphicscale{\twocolumn@sw{0.3}{0.4}}
\def\graphicthreescale{\twocolumn@sw{0.3}{0.4}}

\begin{document}

\title{Dynamic Kibble-Zurek scaling framework \\
for open dissipative many-body systems crossing quantum transitions}

\author{Davide Rossini}
\affiliation{Dipartimento di Fisica dell'Universit\`a di Pisa
        and INFN, Largo Pontecorvo 3, I-56127 Pisa, Italy}

\author{Ettore Vicari} 
\affiliation{Dipartimento di Fisica dell'Universit\`a di Pisa
        and INFN, Largo Pontecorvo 3, I-56127 Pisa, Italy}

\date{\today}

\begin{abstract}
  We study the quantum dynamics of many-body systems, in the presence
  of dissipation due to the interaction with the environment, under
  Kibble-Zurek (KZ) protocols in which one Hamiltonian parameter is
  slowly, and linearly in time, driven across the critical value of a
  zero-temperature quantum transition.  In particular we address
  whether, and under which conditions, open quantum systems can
  develop a universal dynamic scaling regime similar to that emerging
  in closed systems.  We focus on a class of dissipative mechanisms
  whose dynamics can be reliably described through a Lindblad master
  equation governing the time evolution of the system's density
  matrix.  We argue that a dynamic scaling limit exists even in the
  presence of dissipation, whose main features are controlled by the
  universality class of the quantum transition. This requires a
  particular tuning of the dissipative interactions, whose decay rate
  $u$ should scale as $u\sim t_s^{-\kappa}$ with increasing the time
  scale $t_s$ of the KZ protocol, where the exponent $\kappa =
  z/(y_\mu+z)$ depends on the dynamic exponent $z$ and the
  renormalization-group dimension $y_\mu$ of the driving Hamiltonian
  parameter.  Our dynamic scaling arguments are supported by numerical
  results for KZ protocols applied to a one-dimensional fermionic wire
  undergoing a quantum transition in the same universality class of
  the quantum Ising chain, in the presence of dissipative mechanisms
  which include local pumping, decay, and dephasing.
\end{abstract}

\maketitle


\section{Introduction}
\label{intro}

The recent experimental progress in the control and manipulation of
quantum many-body systems has led to great achievements, opening the
door for the realization of quantum simulators~\cite{CZ-12, BDN-12,
  BR-12, AW-12, HTK-12, GAN-14}.  However, the effective isolation of
a quantum system remains a challenge, since interactions with the
environment can have a significant impact in the dynamics, even when
they interact weakly.  It is thus important to understand the effects
of dissipative interactions, irrespectively of their strength.  This
issue is of particular relevance for systems at quantum
transitions~\cite{Sachdev-book}, where the above mentioned effects are
generally relevant, thus tending to suppress the critical quantum
correlations~\cite{NRV-19-dis, RV-19-dis}.

Slow passages through quantum transitions allow us to probe some
universal features of quantum fluctuations in such circumstances.  In
this respect, we mention the Kibble-Zurek (KZ)
problem~\cite{Kibble-76, Zurek-85, ZDZ-05, PG-08, CEGS-12}, related to
the amount of final defects, after slow (quasi adiabatic) passages
through continuous quantum transitions, from the disorder phase to the
order phase. Its scaling predictions have been confirmed by
experiments for various physically interesting systems, see e.g.
Refs.~\cite{Ulm-etal-13, Pyka-etal-13, Lamporesi-etal-13,
  Gong-etal-16, ARBBHC-16, Keesking-etal-19}.  KZ-like protocols have
been largely employed to investigate the critical dynamics of closed
systems, subject to unitary time evolutions only~\cite{Dziarmaga-10,
  PSSV-11}.  The open nature quantum systems, however, may lead to a
departure from the dynamic scaling behavior predicted for the isolated
case~\cite{FFO-07, PSAFS-08, PASFS-09, NVC-15, DRC-16,
  GZYZ-17,KMSFR-17, SVPKD-17, ABRS-18, GTC-19, PSHP-19, FFCQE-20}.  In
particular, it has been observed that slower quenches in open systems,
or subject to noisy controls, may generate an overabundance of defects
when approaching the adiabatic limit in KZ protocols, sometimes named
anti-KZ behavior~\cite{GLDKFS-12}.

Since dissipative mechanisms are expected to give rise to relevant
perturbations at the quantum criticality of closed
systems~\cite{RV-19-dis, NRV-19-dis, YMZ-14} (such as the
temperature), they do not generally preserve the universal dynamic
properties of quantum transitions.  From this point of view, the
above-mentioned anti-KZ behavior should not be considered as
unexpected. Indeed, due to the general relevance of the perturbations
associated with dissipative mechanisms, slower protocols favor the
dissipation effects, in that they give them more time to act.
Therefore, unlike closed systems, the dynamic behaviors arising
from slow changes of the Hamiltonian parameters, across their critical
values, do not anymore develop universal critical features controlled
by the quantum transition of the closed system.  Only an appropriate
tuning of the dissipation strength may give rise to a nontrivial
interplay with the critical unitary dynamics, developing a dynamic
scaling behavior in KZ protocols controlled by the universality class
of the quantum transition.

The issue we address in this paper is whether, and under which
conditions, open dissipative systems may still present a universal
regime controlled by the universality class of the quantum transition
of the closed system. We focus on a class of dissipative mechanisms
whose dynamics can be reliably described through a Lindblad master
equation~\cite{Lindblad-76, GKS-76} governing the time evolution of
the density matrix of the system~\cite{BP-book, RH-book, SBD-16}.  We
argue that, in the presence of weak dissipation, the dynamics of
many-body systems may still develop a scaling behavior under KZ
protocols (i.e., slow changes of one Hamiltonian parameter across its
critical value), thus extending the dynamic KZ scaling of closed
systems~\cite{CEGS-12}. Its main features, in the presence of weak
dissipation, are still controlled by the universality class of the
quantum transition, provided the system-environment interaction
strength is suitably tuned. This allows us to define a dynamic KZ
scaling limit in the presence of dissipation. In particular, we argue
that the decay rate $u$ of the dissipative interactions must scale as
a power law $u\sim t_s^{-\kappa}$ with increasing the time scale $t_s$
of the slow variations of the KZ protocol, where $\kappa<1$ is an
appropriate positive exponent, depending on the universal critical
exponents of the quantum transition.  The suppression of the
dissipation rate is necessary to observe universal dynamic scaling and
it is analogous to that found in other dynamic problems with
dissipation at quantum transitions~\cite{NRV-19-dis, RV-19-dis}.

To check our general framework, we present a numerical analysis of KZ
protocols applied to the fermionic Kitaev wire~\cite{Kitaev-01} across
its quantum transition (belonging to the same universality class of
the one appearing in quantum Ising chains) in the presence of
dissipative mechanisms including local pumping, decay, and dephasing. 
This model can be exactly and fully solved (i.e.~with respect to its
full excitation spectrum) even with a large number of sites,
up to a few thousands, thus enabling up to perform an accurate
numerical investigation of the dynamic KZ scaling behavior put forward.
Our results nicely confirm the emerging of a dynamic
scaling in the limit of slow passages across the quantum transition
and in the presence of weak dissipation.

The paper is organized as follows.  In Sec.~\ref{KZprot} we describe
our dynamic KZ protocol and discuss the Lindblad modelization of a
dissipative system-environment interaction.  In Sec.~\ref{closedsy} we
summarize the main features of the dynamic KZ scaling limit, and the
emerging scaling laws in the limit of large time scale of the KZ
protocol.  In Sec.~\ref{dissint} we extend the dynamic KZ scaling laws
to allow for the presence of dissipation, thus achieving a unique
framework to discuss the interplay between (critical) coherent and
dissipative drivings. Subsequently we introduce the open-system Kitaev
quantum wire, which represents our theoretical laboratory to check the
phenomenological dynamic KZ scaling we put forward
(Sec.~\ref{Kitaevmo}), and present extensive numerical analyses of KZ
protocols for that model at its zero-temperature quantum transition in
the presence of dissipation (Sec.~\ref{numres}). Finally,
Sec.~\ref{conclu} contains a brief summary and some concluding remarks.

\section{Dynamic KZ protocol in the presence of dissipation}
\label{KZprot}

We consider a many-body system presenting a quantum transition driven
by the Hamiltonian parameters.  For simplicity, we assume that the
Hamiltonian $\hat H$ depends on a single relevant parameter
$\mu$, whose variation drives a quantum transition separating two
different quantum phases.  The deviation
\begin{equation}
  \bar{\mu}\equiv \mu - \mu_c
  \label{barmudef}
\end{equation}
quantifies the distance from the critical point, located at
$\mu=\mu_c$.  We also suppose that negative values $\bar{\mu}<0$
correspond to the gapped quantum disordered phase.  Quasi-adiabatic
passages through the quantum transition, slowly varying $\mu$ across
$\mu=\mu_c$, give rise to peculiar out-of-equilibrium phenomena, such
as the one related to the so-called KZ problem~\cite{Zurek-85, ZDZ-05,
  PG-08, Dziarmaga-10, PSSV-11} addressing the formation of defects
when passing through quantum critical points, from the gapped
disordered phase to the ordered phase.

A standard KZ protocol would proceed as follows: {\it i)}~One starts
from the ground state of the many-body system at $\bar\mu_i < 0$, or
alternatively from a statistical state described by the Gibbs
distribution $\propto e^{-\hat H(\bar\mu_i)/k_B T}$ at small
temperature $T$; {\it (ii)}~Then the quantum dynamics is driven by
slow variations of the relevant parameter $\bar\mu$ associated with
the quantum transition, for example linearly as
\begin{equation}
  \bar\mu(t) = t/t_s\,,
  \label{mut}
\end{equation}
up to a value $\bar\mu_f>0$.  The parameter $t_s>0$ denotes the time
scale of the slow variations of the Hamiltonian parameter $\bar\mu$.
The time evolution is unitary, i.e.
\begin{equation}
  {\partial\rho\over \partial t} = -{i\over \hslash}\big[ \hat
    H(\bar\mu),\rho \big]\,,
  \label{heieq}
\end{equation}
where $\rho(t)$ is the density matrix of the many-body system.  Even
in the limit of very slow changes, corresponding to $t_s\to\infty$,
infinite-volume systems cannot satisfy the adiabatic dynamic condition
when passing through the transition point, thus developing
out-of-equilibrium behaviors.  The resulting evolution of the system
is usually investigated by monitoring observables obtained by taking
expectation values at fixed time. For example, in the case of lattice
spin models, one may consider the magnetization, the two-point
function of local operators related to the order parameter, etc \ldots

Here we want to study the effects of weak dissipative mechanisms on
the slow dynamics across the quantum transition. Therefore, beside the
changes of the Hamiltonian parameters, we suppose that the many-body
system is also subject to some interaction with the environment.  The
time dependence of its density matrix $\rho$ can be reasonably
described by the Lindblad master equation~\cite{BP-book}
\begin{equation}
  {\partial\rho\over \partial t} = -{i\over \hslash} \big[ \hat
    H(\bar\mu),\rho \big] + u \,{\mathbb D}[\rho]\,,
  \label{lindblaseq}
\end{equation}
where the first term in the right-hand side provides the coherent
driving, while the second term accounts for the coupling to the
environment, characterized by a global coupling constant $u>0$.

We restrict to homogeneous dissipation mechanisms, preserving
translational invariance. In the case of systems weakly coupled to
Markovian baths, the trace-preserving superoperator can be written as
a sum of local terms, such as~\cite{Lindblad-76, GKS-76}
\begin{eqnarray}
  {\mathbb D}[\rho] & = & \sum_o {\mathbb D}_o[\rho] \,, \label{supop}\\
  {\mathbb D}_o[\rho] & = & \hat L_o \rho \hat L_o^\dagger - \tfrac{1}{2}
  \big( \rho\, \hat L_o^\dagger \hat L_o + \hat L_o^\dagger \hat L_o \rho \big)\,,
  \label{dL}
\end{eqnarray}
where $\hat L_o$ is the Lindblad jump operator associated with the
local system-bath coupling scheme, and $o$ denotes an appropriate
spatial coordinate.  In quantum optical implementations, the
conditions leading to Eqs.~\eqref{lindblaseq}-\eqref{dL} are typically
satisfied~\cite{SBD-16}, therefore this formalism constitutes the
standard choice for theoretical investigations of this kind of
systems.

In the following we analyze the dynamic scaling behavior arising from
dynamic protocols of quantum many-body systems in the presence of weak
dissipation, thus evolving according to Eq.~(\ref{lindblaseq}), when
the parameter $\mu$ is slowly varied across its critical value $\mu_c$
associated with the quantum transition driven by the Hamiltonian,
starting from the gapped disordered phase, analogously to the standard
KZ protocol for closed systems.

\section{Dynamic KZ scaling for closed quantum systems}
\label{closedsy}

Before discussing the effects of dissipation, we recall the main
features of the dynamic scaling behavior developed by many-body
systems unitarily evolving at quantum transitions~\cite{GZHF-10,
  CEGS-12, PRV-18, PRV-18-lo}, and in particular when they are slowly
driven across its quantum transition, according to the KZ protocol
described in Sec.~\ref{KZprot}, cf.~Eqs.~\eqref{mut}-\eqref{heieq}.

\subsection{Homogeneous scaling laws}
\label{homolaws}

At the critical point, the low-energy unitary Hamiltonian dynamics
develops long-distance correlations, characterized by a diverging
length scale $\xi\sim |\bar\mu|^{-\nu}$ [where $\nu=1/y_\mu$ and
  $y_\mu$ is the renormalization-group (RG) dimension of the relevant
  parameter] and the suppression of the gap (energy difference between
the lowest states) $\Delta\sim\xi^{-z}$.  The correlation-length
exponent $\nu$ and the dynamic exponent $z$ are the critical exponents
associated with the universality class of the quantum transition.  The
dynamics at continuous quantum transitions develop homogeneous scaling
laws~\cite{ZDZ-05, CEGS-12, Dziarmaga-05, GZHF-10, PRV-18, PRV-18-lo,
  Biroli-15, CC-16, v-18, NRV-19-wf, RV-19-de}, even in the presence
of interactions with an environment~\cite{NRV-19-dis, RV-19-dis,
  YMZ-14, RV-20}.

For example, in the case of instantaneous quenches of closed systems,
arising from the instantaneous variation of the Hamiltonian parameter
from $\bar\mu_i$ to $\bar\mu$, starting from the ground state at
$\bar\mu_i$, the evolution of a generic observable $B$, such as the
expectation value of a local operator $\hat{B}$ (assuming translation
invariance), satisfies the homogeneous scaling relation~\cite{PRV-18}
\begin{eqnarray}
  B(\bar\mu_i,\bar\mu,t,L) & \equiv & \langle \Psi(t) | \hat{B} |
  \Psi(t) \rangle \nonumber \\ & \approx & b^{-y_B} \, {\cal
    B}(\bar\mu_i b^{y_\mu}, \bar\mu b^{y_\mu}, t b^{-z}, L/b)\,.
  \label{squsca}  
\end{eqnarray}
Here $|\Psi(t)\rangle$ indicates the quantum many-body state after
the quench, $b$ is an arbitrary positive parameter, $y_{B}$ is the RG
dimension of the operator $\hat{B}$, $L$ is the size of the system,
and ${\cal B}$ is a universal scaling function apart from
normalizations.  Eq.~(\ref{squsca}) is expected to provide the
asymptotic power-law behavior in the large-$b$ limit.

The KZ protocol focuses on the opposite quasi-adiabatic regime, where
the driving parameter $\bar\mu$ is slowly varied across the quantum
transition, starting from the ground state at a given $\bar\mu_i<0$
and then changing $\bar\mu$ linearly in time, as in Eq.~\eqref{mut}
(thus the initial condition $\bar\mu_i$ corresponds to the initial
time $t_i = t_s \bar\mu_i$).  A phenomenological scaling theory is
obtained by assuming the homogeneous scaling law
\begin{subequations}
\begin{equation}
  B(\bar\mu_i,t,t_s,L) \approx
  b^{-y_B} \, {\cal B}(\bar\mu_i b^{y_\mu}, \bar\mu(t) b^{y_\mu}, t b^{-z}, L/b)\,,
  \label{sdynscab}
\end{equation}
where, again, $b$ is an arbitrary positive parameter.  Analogous
scaling equations can be written down for the fixed-time correlations
$G_{AB}$ of two local operators $\hat{A}$ and $\hat{B}$ at a distance
$x$.  Assuming translation invariance,
\begin{eqnarray}
  G_{AB}(x,\bar\mu_i,t,t_s,L) & \!\!\! \equiv \!\!\! & \langle \Psi(t)| \hat{A}(x_0)
  \hat{B}(x_0+x)|\Psi(t) \rangle \label{g12sca} \\
  & \! \approx \!\! & b^{-\varphi} {\cal G}(x/b,\bar\mu_i b^{y_\mu}, \bar\mu(t)
  b^{y_\mu}, t b^{-z}\!, L/b), \nonumber
\end{eqnarray}
\end{subequations}
where $\varphi=y_A+y_B$ and $y_A,y_B$ are the RG dimensions of the
operators $\hat{A}$ and $\hat{B}$, respectively.

The dynamic KZ scaling framework can be extended to situations where
the initial condition is given by a Gibbs ensemble at temperature $T$,
by adding a further dependence on the product $T b^z$ in the KZ
scaling functions of Eqs.~(\ref{sdynscab}) and~(\ref{g12sca}).

\subsection{Dynamic scaling in the infinite-volume limit}
\label{dyninfL}

We now concentrate on KZ protocols.  To derive a dynamic scaling
theory for infinite-volume systems, it is possible to exploit the
arbitrariness of the scale parameter $b$ in the general homogeneous
power laws (\ref{sdynscab}) and (\ref{g12sca}). To this purpose we set
\begin{equation}
  b = \lambda \equiv t_s^{1\over y_\mu+z}\,,
  \label{sbla}
\end{equation}
where $\lambda$ is the length scale associated with the KZ protocol,
and take the limit $L/\lambda\to\infty$ (corresponding to taking the
so-called thermodynamic limit).  This leads to the dynamic KZ scaling
ansatz
\begin{subequations}
\begin{eqnarray}
  B(\bar\mu_i,t,t_s) & \approx & \lambda^{-y_B}\,
  {\cal B}_i(\bar\mu_i \lambda^{y_\mu}, \tau )\,,
  \label{kzB}\\
  G_{AB}(x,\bar\mu_i,t,t_s) & \approx & \lambda^{-\varphi} \,{\cal
    G}_i(x/\lambda,\bar\mu_i \lambda^{y_\mu}, \tau)\,,
  \label{kzG}
\end{eqnarray}
\end{subequations}
where $\tau$ is the rescaled time:
\begin{equation}
  \tau \equiv {t / t_s^{\kappa}}\,,\qquad \kappa= {z\over y_\mu+z}\,.
  \label{tauvar}
\end{equation}
The dynamic KZ scaling limit, where the above asymptotic behaviors
apply, is obtained by taking $t_s\to\infty$ keeping the arguments of
the dynamic scaling functions ${\cal B}_i$ and ${\cal G}_i$ fixed.
Actually, introducing a time scaling variable related to initial time
of the KZ protocol,
\begin{equation}
\tau_i \equiv {t_i / t_s^{\kappa}}\,,\qquad t_i = \bar\mu_i\, t_s\,,
\label{tauidef}
\end{equation}
we may rewrite the scaling Eqs.~(\ref{kzB}) and (\ref{kzG}) as
\begin{subequations}
\begin{eqnarray}
  B(\bar\mu_i,t,t_s) & \approx & \lambda^{-y_B}\, \widetilde {\cal B}_i(\tau_i, \tau )\,,
  \label{kzB2}\\
  G_{AB}(x,\bar\mu_i,t,t_s) & \approx & \lambda^{-\varphi} \,
  \widetilde {\cal G}_i(x/\lambda,\tau_i, \tau)\,.
  \label{kzG2}
\end{eqnarray}
\end{subequations}
Note that the scaling functions ${\cal B}_i, \, {\cal G}_i$ and
$\widetilde {\cal B}_i, \, \widetilde {\cal G}_i$ in
Eqs.~(\ref{kzB})-(\ref{kzG}) and (\ref{kzB2})-(\ref{kzG2}) do not
coincide, but are trivially related by the change of scaling
variables.

Since the KZ protocol starts from $\bar\mu_i<0$ corresponding to the
gapped phase, whose gap decreases as $\Delta\sim \xi^{-z}$ and the
ground-state length scale $\xi$ diverges only at the critical point
$\bar\mu=0$, the emerging dynamic KZ scaling should be independent of
the actual finite value of $\bar\mu_i < 0$, if this is kept fixed in
the dynamic KZ scaling limit.  This is essentially due to the fact
that, in a gapped phase, the evolution arising from slow changes of
the parameters is essentially adiabatic, from $\bar\mu_i$ to the
relevant scaling interval $\delta_\mu$ around $\bar\mu=0$, which
effectively decreases as
\begin{equation}
  \delta_{\bar\mu} \sim t_s^{-1+\kappa} \;\to\; 0
  \label{deltamu}
\end{equation}
in the dynamic KZ scaling limit.  Therefore, when increasing $t_s$,
keeping $\mu_i<0$ constant and finite, the dynamic KZ scaling must be
independent of $\bar\mu_i$, corresponding to the $\tau_i \to -\infty$
limit of the relations~(\ref{kzB2}) and~(\ref{kzG2}).  Therefore this
leads to the dynamic scaling ansatz
\begin{subequations}
\begin{eqnarray}
  B(\bar\mu_i,t,t_s)&\approx& \lambda^{-y_B}\, {\cal B}_\infty (\tau)\,,
  \label{sinflimB}\\
  G_{AB}(x,\bar\mu_i,t,t_s) &\approx& \lambda^{-\varphi} \,
  {\cal G}_\infty (x/\lambda,\tau)\,.
  \label{sinflimG}
\end{eqnarray}
\end{subequations}

The dynamic scaling functions introduced above are expected to be
universal with respect to changes of the microscopic details of the
Hamiltonian within the given universality class.  Of course, like any
scaling function at quantum transitions, such a universality holds
apart from a multiplicative overall constant and normalizations of the
scaling variables.  The approach to the asymptotic dynamic scaling
behavior is expected to be generally characterized by power-law
suppressed corrections.

We finally mention that the so-called KZ problem genuinely addresses
the formation of defects when slowly crossing the quantum transition,
from the disordered to the ordered phase.  The above scaling arguments
in the dynamic KZ limit (see Refs.~\cite{Zurek-85, PSSV-11, CEGS-12})
lead to the expectation that the density of defects arising after
crossing the transition scales as the inverse scaling volume
$\lambda^{-d}$, cf. Eq.~(\ref{sbla}), that is
\begin{equation}
  \rho_{\rm defects} \sim \lambda^{-d} = t_s^{-{d\over y_\mu+z}}\,.
  \label{rhodef}
\end{equation}
This scaling behavior has been verified in experiments, see, e.g.,
Refs.\cite{Ulm-etal-13, Pyka-etal-13, Lamporesi-etal-13, Gong-etal-16,
  Keesking-etal-19, DRGA-99, MMARK-06, GLDKFS-12}.

\subsection{Dynamic finite-size scaling}
\label{dynfss}

The scaling Eqs.~(\ref{sdynscab}) and (\ref{g12sca}) also allow us to
derive dynamic finite-size scaling (FSS) relations, which are valid
far from the thermodynamic limit, and which extend those predicted by
the FSS theory for systems at equilibrium~\cite{SGCS-97, CPV-14,
  CNPV-14}.  For example, by setting $b=L$ in Eq.~(\ref{g12sca}), we
obtain
\begin{equation}
  G_{AB}(x,\bar\mu_i,t,t_s,L) \approx  L^{-\varphi}\, 
  {\cal G}_L(x/L,\bar\mu_i L^{y_\mu} , \bar\mu(t) L^{y_\mu}, t L^{-z} ).
  \label{kzfssG}
\end{equation}
This dynamic FSS behavior is expected to be obtained by taking
$L\to\infty$, while keeping the arguments of the scaling function
${\cal G}_L$ fixed.  One may introduce more convenient scaling
variables, which are combinations of those entering
Eq.~(\ref{kzfssG}).  For example, one can write it as
\begin{equation}
  G_{AB}(x,\bar\mu_i,t,t_s,L) \approx   
  L^{-\varphi}\, {\cal G}_L(x/L,\tau_i, \tau,\upsilon)\,,
  \label{kzfssG2} 
\end{equation}
where 
\begin{equation}
  \upsilon \equiv t_s /L^{y_\mu+z} \label{iupsvar}
\end{equation}
and $\tau,\,\tau_i$ are defined in Eqs.~(\ref{tauvar}) and
(\ref{tauidef}), respectively.

Assuming again that the KZ protocol starts from the gapped disordered
phase and the initial $\bar \mu_i < 0$ is kept fixed in the dynamic
scaling limit, the same dynamic FSS is expected to hold, irrespective
of the value of $\bar \mu_i$.  Thus, the dynamic FSS in
Eq.~\eqref{kzfssG2} simplifies into
\begin{equation}
  G_{AB}(x,\bar\mu_i,t,t_s,L) \approx   
  L^{-\varphi} \, {\cal G}_{L,\infty} (x/L,\tau,\upsilon)\,,
  \label{kzfssG2mufixed} 
\end{equation}
Indeed, with increasing $L$, the dynamic FSS occurs within a smaller
and smaller interval of values of $|\mu|$ around $\bar\mu=0$: since
the time interval of the dynamic process scales as $t_{\rm sca}\sim
t_s^\kappa$, the relevant interval of values of $|\bar\mu|$ shrinks as
$t_{\rm sca}/t_s\sim L^{-y_\mu}$.

Note that, in the limit $\upsilon\to\infty$, the evolution as a
function of $\bar\mu(t)=t/t_s$ corresponds to an adiabatic
dynamics. Indeed, since the finite size $L$ guarantees the presence of
a gap between the lowest states, one may adiabatically cross the
critical point in the limit $\upsilon\to\infty$, passing through the
ground states of the finite-size system for $\bar\mu(t)$. The
adiabatic evolution across the transition point is prevented only when
$L\to\infty$ (before the limit $t_s\to\infty$), i.e., when the time
scale of the critical correlations diverges, since
$\tau_{\rm cr} \sim \Delta^{-1}\sim L^z$.

\section{Dynamic KZ scaling for open quantum systems}
\label{dissint}

\subsection{Dynamic scaling allowing for dissipation}
\label{phedynsca}

In this section, we extend the dynamic scaling theory outlined in
Sec.~\ref{closedsy} to systems subject to dissipative interactions
with the environment, so that the time dependence of the density
matrix $\rho$ is described by the Lindblad master equation
(\ref{lindblaseq}). Namely, we assume that the quantum evolution
arising from the KZ protocol occurs in the presence of dissipation
with the effective coupling $u>0$, thus being ruled by
Eq.~(\ref{lindblaseq}).  The resulting dynamic KZ scaling framework
will provide a unique framework to discuss the interplay between
(critical) coherent and dissipative drivings.

The dynamic behavior in the presence of weak dissipation has been
addressed within a phenomenological dynamic scaling theory in
Refs.~\cite{NRV-19-dis, RV-19-dis}, extending the dynamic scaling
scenario holding for closed systems.  This has been obtained by adding
a further dependence associated with the dissipation parameter $u$ in
the dynamic scaling relations~\eqref{sdynscab} and~\eqref{g12sca},
through a power law $u b^z$, where the dynamic exponent $z$ ensures
the substantial balance (i.e., competition) with the critical coherent
driving.  We recall that this hypothesis has been put forward after
noting that the parameter $u$ of the dissipator in
Eq.~\eqref{lindblaseq} plays the role of a decay rate, i.e., of an
inverse relaxation time, of the associated dissipative
process~\cite{BP-book}.  Thus, to observe a nontrivial competition
between critical coherent dynamics and dissipation, the dissipative
coupling must be comparable to the gap of the critical Hamiltonian,
therefore its scaling variable must be controlled by the dynamic
exponent $z$.

Following the above reasoning, we conjecture that KZ protocols in the
presence of dissipation develop homogeneous laws, such as
\begin{subequations}
\begin{eqnarray}
  &&B(\bar\mu_i,t,t_s,u,L) \equiv {\rm Tr}[\rho(t)\hat{B}] \qquad \nonumber \\
  && \quad \approx b^{-y_B}\,
  {\cal B}(\bar\mu_i b^{y_\mu}, \bar\mu(t) b^{y_\mu}, t b^{-z}, L/b, u b^z) \,,\qquad
  \label{sdynscabdis}
\end{eqnarray}
and
\begin{eqnarray}
  &&G_{AB}(x,\bar\mu_i,t,t_s,u,L) \equiv {\rm Tr}[\rho(t)\hat{A} \hat{B}]
    \qquad \nonumber \\
  &&\quad \approx b^{-\varphi} \,{\cal G}(x/b,\bar\mu_i b^{y_\mu},
  \bar\mu(t) b^{y_\mu}, t b^{-z}, L/b, u b^z)\,, \qquad
  \label{g12scadis}
\end{eqnarray}
\end{subequations}
similar to those in Eqs.~\eqref{sdynscab} and~\eqref{g12sca},
but with one additional scaling variable associated with $u$.

\subsection{Dynamic scaling in the infinite-volume limit}

Analogously to the dynamics of closed systems, it is possible to
derive scaling laws in the thermodynamic limit, by fixing $b$ as in
Eq.~(\ref{sbla}) and taking $L/\lambda\to\infty$.  One can easily show
that Eqs.~(\ref{sdynscabdis}) and~\eqref{g12scadis} imply the dynamic
KZ scaling ansatz
\begin{subequations}
  \begin{eqnarray}
    B(\bar\mu_i,t,t_s,u) & \approx & 
    \lambda^{-y_B}\, {\cal B}_i(\tau_i,\tau,\gamma)\,,
    \label{inflimdis}\\
    G_{AB}(x,\bar\mu_i,t,t_s,u) & \approx &
    \lambda^{-\varphi} \, {\cal G}_i(x/\lambda,\tau_i,\tau,\gamma)\,,
    \label{inflimdisG}
\end{eqnarray}
\end{subequations}  
where we introduced the scaling variable $\gamma$ associated with the
dissipation parameter,
\begin{equation}
 \gamma = u\, t_s^\kappa\,,\qquad \kappa= {z\over y_\mu+z}\,.
  \label{hatgdef}
\end{equation}
The above scaling laws are expected to provide the asymptotic behavior
in the $t_s\to\infty$ limit while keeping the scaling variables fixed,
including $\gamma$.

Note that, like for closed systems, the large-$t_s$ limit of KZ
protocols starting from finite and fixed $\bar\mu_i<0$ should
correspond to the limit $\tau_i\to -\infty$ in the right-hand side of
Eqs.~(\ref{inflimdis}) and (\ref{inflimdisG}).  
Indeed, the dissipation with coupling strength $u\sim \lambda^{-z}$ is
not expected to play any relevant role at finite $\bar\mu_i<0$, where
the gap is $\Delta = O(1)$, while it should compete with the unitary
evolution only very close to $\bar\mu=0$ where $u\sim \Delta \sim \lambda^{-z}$.
Therefore, under such conditions we expect the scaling behavior
\begin{subequations}
\begin{eqnarray}
  B(\bar\mu_i,t,t_s,u) & \approx &   
  \lambda^{-y_B}\, {\cal B}_\infty (\tau,\gamma)\,,
  \label{sinflimdis}\\
  G_{AB}(x,\bar\mu_i,t,t_s,u) & \approx &
  \lambda^{-\varphi} \, {\cal G}_\infty (x/\lambda,\tau,\gamma)\,.
  \label{sinflimdisG}
\end{eqnarray}
\end{subequations}

We mention that, in the above KZ scaling limit allowing for
dissipation, the scaling law associated with the number of defects,
cf.~Eq.~(\ref{rhodef}), should be replaced with
\begin{equation}
  \rho_{\rm defects} \approx \lambda^{-d} \,{\cal D}(\gamma) = 
  t_s^{-{d\over y_\mu+z}}\,{\cal D}(\gamma)\,,
  \label{rhodefdis}
\end{equation}
where the dependence on the dissipative coupling $u$ enters the
scaling function ${\cal D}$ through the scaling variable $\gamma$.  Of
course, one must recover the scaling law (\ref{rhodef}) for $\gamma=0$.

\subsection{Dynamic finite-size scaling}
\label{fssdis}

The dynamic FSS behavior can be obtained by setting $b=L$ in
Eqs.~(\ref{sdynscabdis}) and (\ref{g12scadis}), thus extending the
results contained in Sec.~\ref{dynfss} to allow for the dissipation
term of the Lindblad equation. Namely,
\begin{equation}
  G_{AB}(x,\bar\mu_i,t,t_s,u,L) \approx
  L^{-\varphi} \, {\cal G}_L(x/L,\tau_i, \tau,\upsilon,\gamma_L)\,,
  \label{kzfssG2dis} 
\end{equation}
where 
\begin{equation}
  \gamma_L = u \,L^z\,.\label{gammadef}
\end{equation}
The above scaling law can be obtained in the $L\to\infty$ limit
while keeping the scaling variables fixed, including $\gamma_L$.

Moreover, assuming again that the quantum phase for $\bar\mu<0$ is
gapped, with $\Delta\sim\xi^{-z}$ and the ground-state length scale
$\xi$ diverges only at the critical point $\bar\mu=0$, KZ protocols
associated with any finite initial $\bar\mu_i<0$ develop the same
dynamic FSS independently of their actual values. Thus, the dynamic
FSS can be written as
\begin{equation}
  G_{AB}(x,\bar\mu_i,t,t_s,u,L) \approx   
  L^{-\varphi}\, {\cal G}_{L,\infty} (x/L,\tau,\upsilon,\gamma_L)\,.
  \label{kzfssG3} 
\end{equation}
Indeed, similarly to the infinite-volume case, the dissipation with
coupling strength $u\sim L^{-z}$ is not expected to play any relevant
role at finite $\bar\mu_i<0$, where the gap is $\Delta = O(1)$, while
it should compete with the unitary evolution only very close to the
critical point where $u\sim \Delta \sim L^{-z}$.  Like for closed
systems, the dynamic scaling limit thus involves smaller and smaller
intervals of values of $|\bar\mu|$ around $\bar\mu=0$ with increasing
$L$: since the time interval of the dynamic process scales as
$t_{\rm sca}\sim t_s^\kappa$, the relevant interval of values of $|\mu|$
shrinks as $t_{\rm sca}/t_s\sim L^{-y_\mu}$.

\section{KZ protocols for the Kitaev quantum wire subject to dissipation}
\label{Kitaevmo}

To verify the dynamic KZ scaling laws put forward in the previous
sections, and in particular in Sec.~\ref{dissint}, we consider a
Kitaev quantum wire defined by the Hamiltonian~\cite{Kitaev-01}
\begin{equation}
 \hat H_{\rm K} = - t \sum_{j=1}^L \big( \hat c_j^\dagger \hat
  c_{j+1} + \delta \, \hat c_j^\dagger \hat c_{j+1}^\dagger+{\rm h.c.}
  \big) - \mu \sum_{j=1}^L \hat n_j \,,
  \label{kitaev2}
\end{equation}
where $\hat c_j$ is the fermionic annihilation operator on the $j$th
site of the chain, $\hat n_j\equiv \hat c_j^\dagger \hat c_j$ is the
density operator, and $\delta>0$.  We set $\hslash =1$, and $t=1$ as
the energy scale. Moreover we fix $\delta=1$.  We consider
antiperiodic boundary conditions, $\hat c_{L+1} = - \hat c_1$, and
even $L$ for computational convenience.

The fermionic system described by the Hamiltonian~(\ref{kitaev2})
undergoes a continuous quantum transition at $\mu=\mu_c = -2$,
independently of $\delta$, belonging to the same universality class of
that of the quantum Ising chain (when $\delta>0$), i.e. the two-dimensional Ising
universality class~\cite{Kitaev-01,Sachdev-book}, characterized by the
length-scale critical exponent $\nu=1$, related to the RG dimension
$y_\mu = 1/\nu=1$ of the Hamiltonian parameter $\mu$ (more precisely
of the difference $\bar{\mu} \equiv \mu-\mu_c$).  The dynamic exponent
associated with the unitary quantum dynamics is $z=1$.  Moreover, the
RG dimension of the fermionic operators $\hat c_j$ and $\hat
c^\dagger_j$ is $y_{\hat c} =y_{\hat c^\dagger} = 1/2$, and that of
the density operator $\hat n_j$ is $y_{\hat n} =
1$~\cite{Sachdev-book}.  Details on the correspondence with the
quantum Ising chain are provided in App.~\ref{app:KitaevIsing}.

We focus on the dynamic behavior of the Fermi lattice
gas~\eqref{kitaev2} close to its quantum transition, in the presence
of homogeneous dissipation mechanisms following the Lindblad equation
(\ref{lindblaseq}). The dissipator ${\mathbb D}[\rho]$ is defined as a
sum of local (single-site) terms of the form
\begin{equation}
  {\mathbb D}_j[\rho] = \hat L_j \rho \hat L_j^\dagger - \tfrac{1}{2}
  \big( \rho\, \hat L_j^\dagger \hat L_j + \hat L_j^\dagger \hat L_j
  \rho \big)\,,
  \label{dLj}
\end{equation}
where $\hat L_j$ denotes the Lindblad jump operator associated with
the system-bath coupling scheme, and the index $j$ corresponds to a
lattice site [thus replacing the index $o$ in
  Eqs.~\eqref{supop},\eqref{dL}].  The onsite Lindblad operators $\hat
L_j$ describe the coupling of each site with an independent bath. We
consider dissipation mechanisms associated with either particle losses
(l), pumping (p), or dephasing (d), respectively~\cite{HC-13,
  KMSFR-17, NRV-19-dis,Davies-70, Evans-77, SW-10,Nigro-19}:
\begin{equation}
  \hat L_{{\rm l},j} = \hat c_j \,, \qquad \hat L_{{\rm p},j} = \hat
  c_j^\dagger \,, \qquad \hat L_{{\rm d},j} = \hat n_j\,.
  \label{loppe}
\end{equation}
The choice of such dissipators turns out to be particularly convenient
for the numerical analysis, allowing us to
scale the difficulty of the problem linearly with $L$ and thus to
obtain results for the Kitaev wire with thousands of sites (see
Sec.~\ref{numres}). This is important, in view of the necessity
to perform adequate numerical checks of a new scaling theory lying
on phenomenological grounds.

The KZ protocol that we consider starts from the ground state of $\hat
H_K$ for a generic $\bar{\mu}_i<0$, where the system is gapped,
$\Delta = |\bar{\mu}| + O(L^{-2})$, while $\Delta\sim L^{-1}$ at
$\bar\mu=0$ (see App.~\ref{app:KitaevIsing}).  Then the system evolves
according to Eq.~(\ref{lindblaseq}) with a time dependent parameter
$\bar\mu(t)=t/t_s$, starting from $t_i<0$ such that
$\bar\mu_i=t_i/t_s$.  To characterize the dynamic properties of the
evolution described by the Lindblad equation, and in particular the
corresponding asymptotic large-time behavior, we consider the
fixed-time correlations
\begin{subequations}
\begin{eqnarray}
P(x,t) & \!\! = \!\! & {\rm Tr}[\rho(t)\,(\hat c_j^\dagger 
\hat c_{j+x}^\dagger +
    \hat c_{j+x} \hat c_{j})],\label{ptf}\\ 
C(x,t) & \!\! = \!\! & {\rm Tr}[\rho(t)\, (\hat c_j^\dagger \hat c_{j+x} 
+ \hat
    c_{j+x}^\dagger \hat c_{j})],\label{gtf}\\ 
G(x,t) & \!\! = \!\! & {\rm Tr}[\rho(t)\, \hat n_j \hat n_{j+x}] \! - \! 
{\rm Tr}[\rho(t)\, \hat n_j] \, {\rm Tr}[\rho(t)\, \hat n_{j+x}],
\qquad \; \label{gntf}
\end{eqnarray}
\end{subequations}
where $j,x \in [1,L/2]$. 

The dynamic KZ scaling of the above correlation functions is expected
to be given by the general scaling laws reported for the generic
two-point function $G_{AB}$ in Sec.~\ref{dissint}, taking into account
that $\varphi=1$ for the correlations $P$ and $C$ (since $y_{\hat
  c}=y_{\hat c^\dagger} = 1/2$), while $\varphi=2$ for $G$ (since
$y_{\hat n}=1$).  This scaling scenario should hold for all the
considered dissipation mechanisms, cf. Eq.~\eqref{loppe}.  Of course,
the corresponding scaling functions are expected to differ.

\section{Numerical results}
\label{numres}

We now present the results of a series of numerical computations
we have performed on the Kitaev quantum wire. As stated above,
this model is amenable to a direct solvability for systems
with $O(10^3)$ sites, thus representing the
ideal playground for open quantum lattice problems, given the
remarkable difficulty to simulate the dynamics of interacting
many-body quantum systems coupled to an external bath.

For the specific choice of dissipators in Eq.~\eqref{loppe}, the
exponential complexity of the Kitaev chain can be semi-analytically
reduced to a polynomial one~\cite{Eisler-11, HC-13, KMSFR-17,
  NRV-19-dis}.  In particular, in the presence of particle losses (l)
or pumping (p) and for translationally invariant systems, the
driven-dissipative quantum dynamics ruled by the master
equation~\eqref{lindblaseq} can be exactly solved by decoupling in
Fourier space the various sectors with different momenta,
analogously to fermionic Gaussian Hamiltonian models.  Similar
strategies can be adopted for more general inhomogeneous
(disordered) lossy dynamics, provided the Liouvillian operator remains
quadratic in the creation and annihilation operators for fermions.
On the other hand, although the quantum dynamics with a dephasing (d)
mechanism cannot be simply obtained, two-point observables are still
fully captured by a set of coupled linear differential equations,
whose number increases linearly with the number of sites $L$ (see,
e.g., the appendix in Ref.~\cite{NRV-19-dis} for details).  The latter
can be integrated, e.g., with a standard fourth-order Runge-Kutta
method.

\begin{figure}[!t]
  \includegraphics[width=0.95\columnwidth]{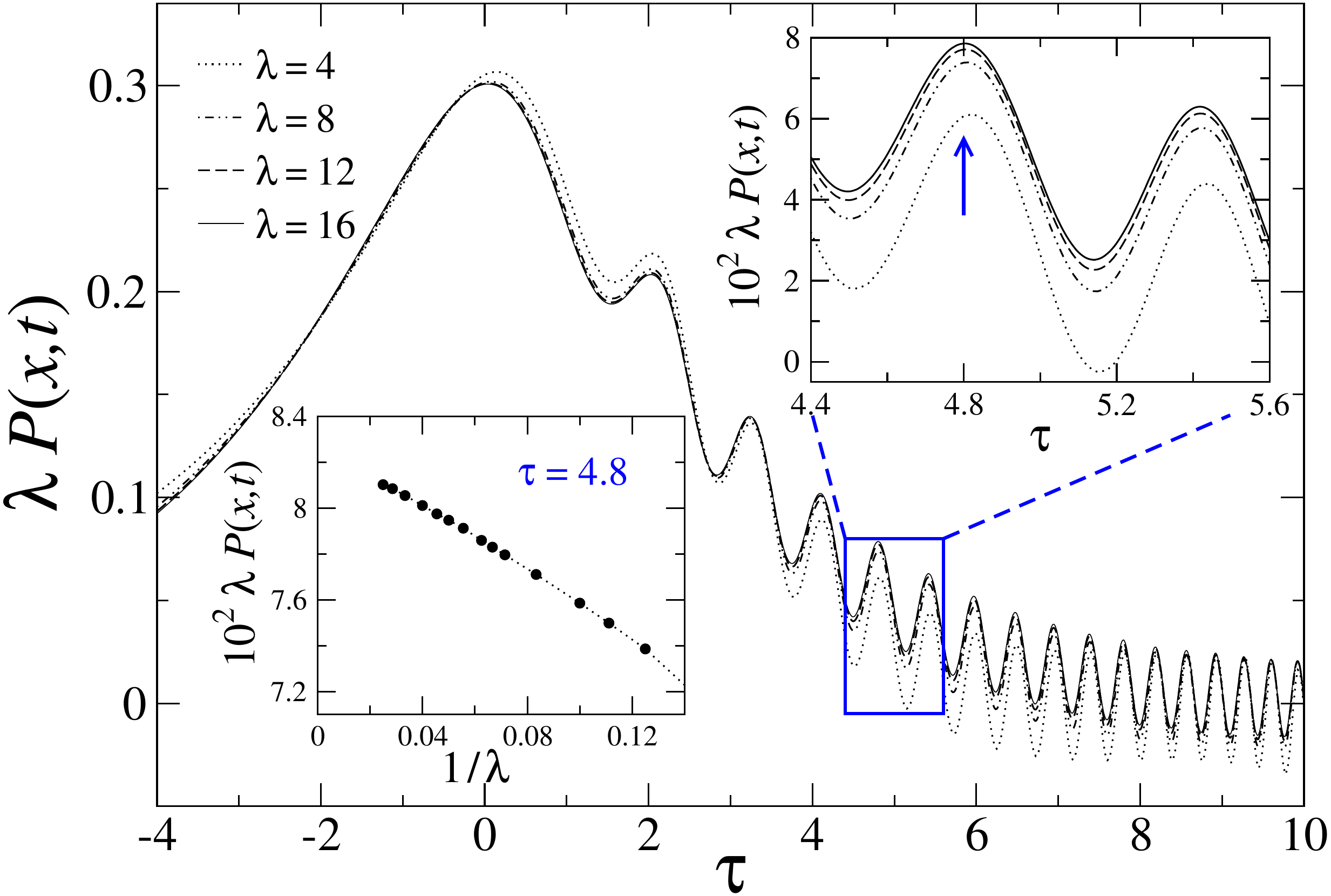}
  \caption{Rescaled correlation $\lambda \, P(x,t)$, fixing $x/\lambda
    = 1$, for the unitary dynamics of the Kitaev quantum wire in the
    thermodynamic limit, as a function of the scaling time variable
    $\tau$. Here we fix the scaling variable associated with the
    initial time, $\tau_i = -10$.  Lines with different styles are for
    various values of the length scale $\lambda$, from $4$ to $16$, as
    indicated in the legend.  The upper right inset shows a
    magnification of the data for $4.4 < \tau < 5.6$, while the lower
    left inset displays rescaled correlations as a function of
    $1/\lambda$, for fixed $\tau = 4.8$ (arrow in the upper inset),
    supporting an $O(\lambda^{-1})$ approach to the asymptotic KZ
    scaling limit.  Analogous results are obtained for other values
    of the scaling variables $x/\lambda$ and $\tau_i$, and for
    the correlations $C(x,t)$ and $G(x,t)$.}
\label{fig:P_TL_nodiss_tau}
\end{figure}

\subsection{Dynamic KZ scaling in the infinite-volume limit}

We first discuss systems in the thermodynamic limit.  To ensure that
finite-size corrections are negligible on the scale of all the
numerics presented below for the dynamic scaling, we have carefully
checked that (in all cases treated herewith) systems of size $L =
2^{12} = 4096$ allow to simulate KZ protocols with a length scale
$\lambda$ up to $O(10^2)$.

\subsubsection{Unitary KZ dynamics}
\label{sec:TLunitay}

\begin{figure}[!t]
  \includegraphics[width=0.95\columnwidth]{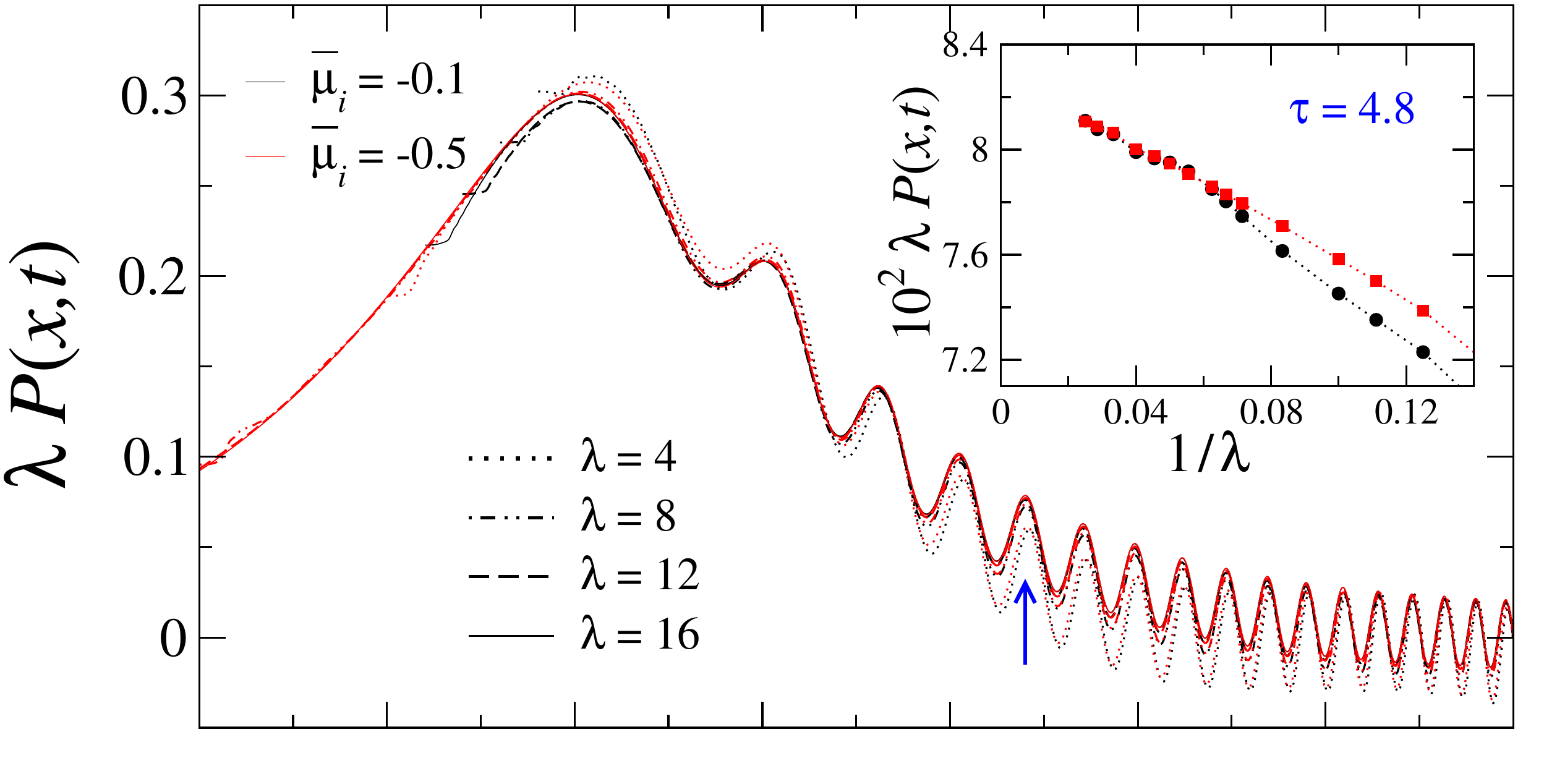}
  \includegraphics[width=0.95\columnwidth]{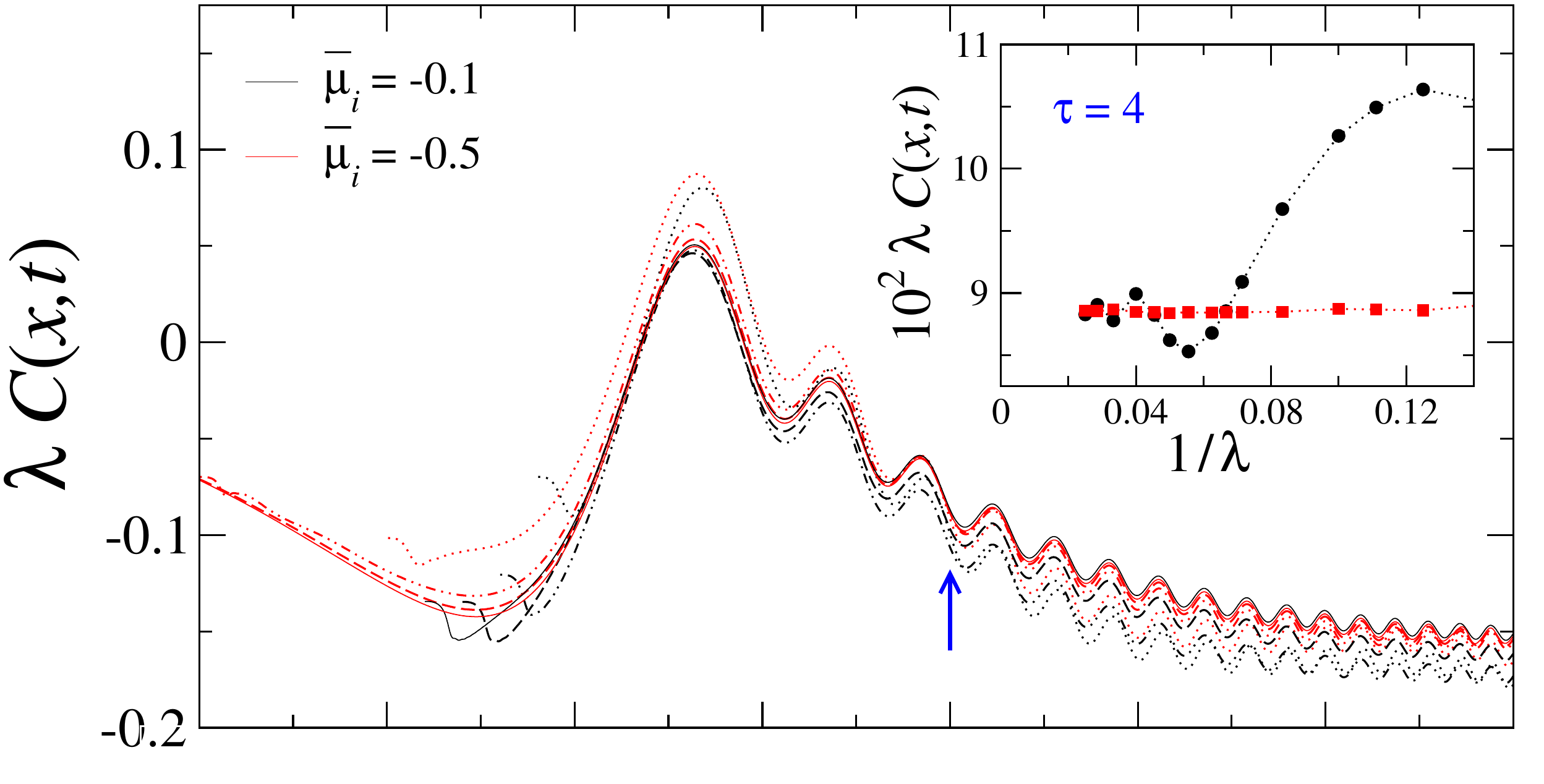}
  \includegraphics[width=0.95\columnwidth]{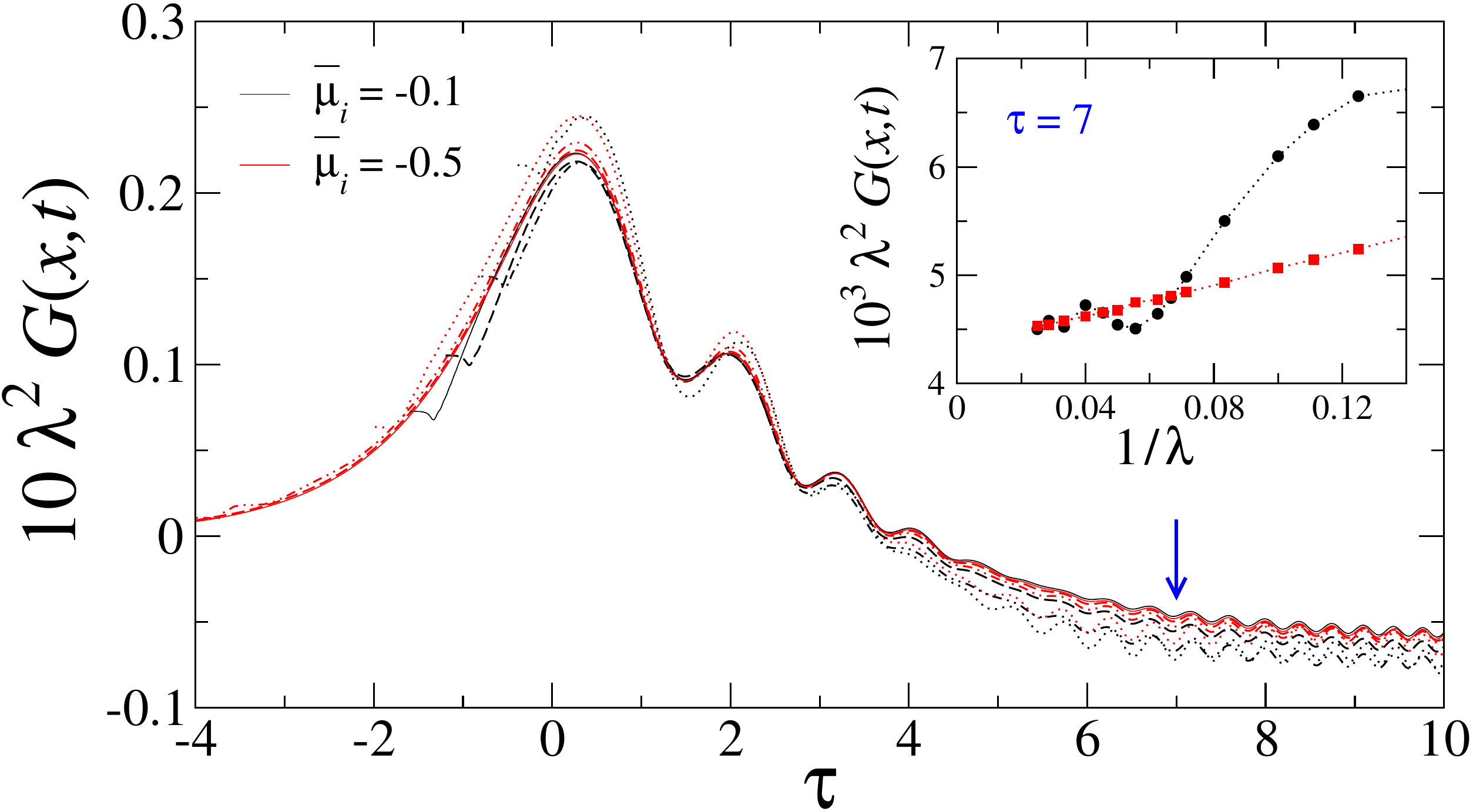}
  \caption{Rescaled correlations $\lambda \, P(x,t)$ (upper panel),
    $\lambda \, C(x,t)$ (central panel), and $\lambda^2 \, G(x,t)$
    (lower panel), at fixed $x/\lambda = 1$ (results for other values
    of $x/\lambda$ show analogous behaviors), for the unitary dynamics
    of the Kitaev quantum wire in the thermodynamic limit, as a
    function of the scaling variable $\tau$.  Different line styles
    stand for various values of the length scale $\lambda$, from $4$
    to $16$, analogously to Fig.~\ref{fig:P_TL_nodiss_tau} (see
    legend).  Data belonging to one of the two color sets correspond
    to a given initial Hamiltonian parameter $\bar \mu_i<0$, which is
    kept fixed and equal to either $\bar \mu_i = -0.1$ (black circles)
    or $\bar \mu_i = -0.5$ (red squares).  The insets in the three
    panels display rescaled correlations as a function of $1/\lambda$,
    for both cases of $\bar \mu_i$ presented in the main frames, at
    the $\tau$ value indicated by the blue arrow.}
\label{fig:PCG_TL_nodiss_mu}
\end{figure}

\begin{figure*}[!t]
  \includegraphics[width=0.95\textwidth]{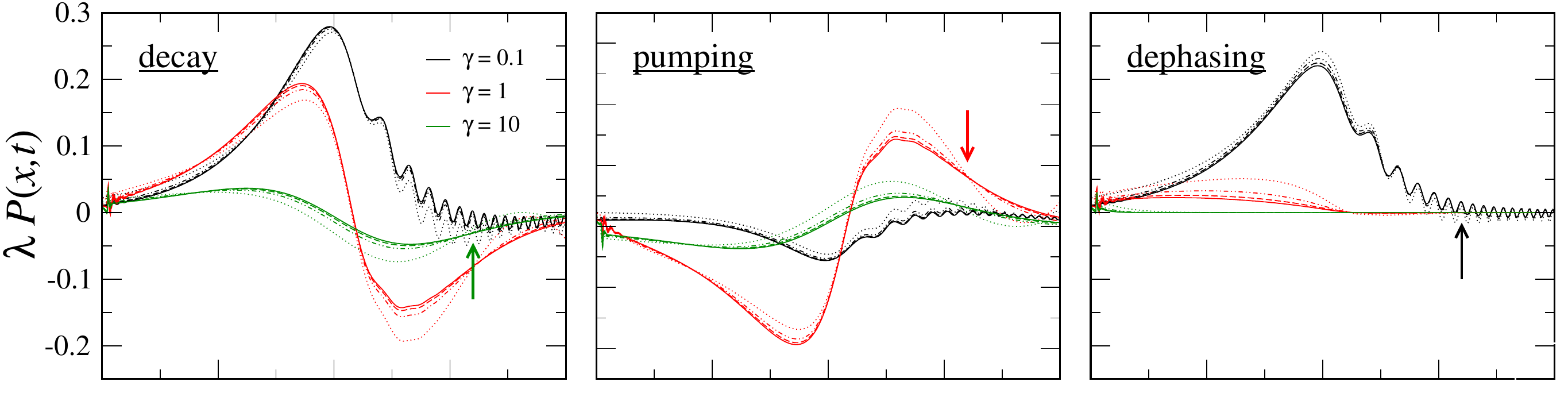}
  \includegraphics[width=0.95\textwidth]{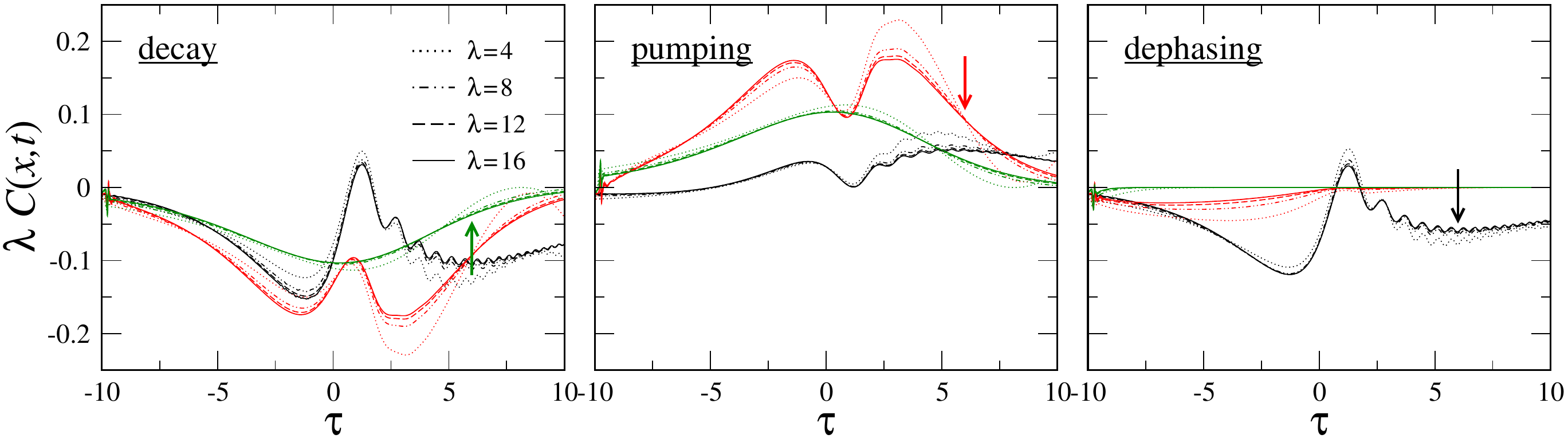}
  \includegraphics[width=0.95\textwidth]{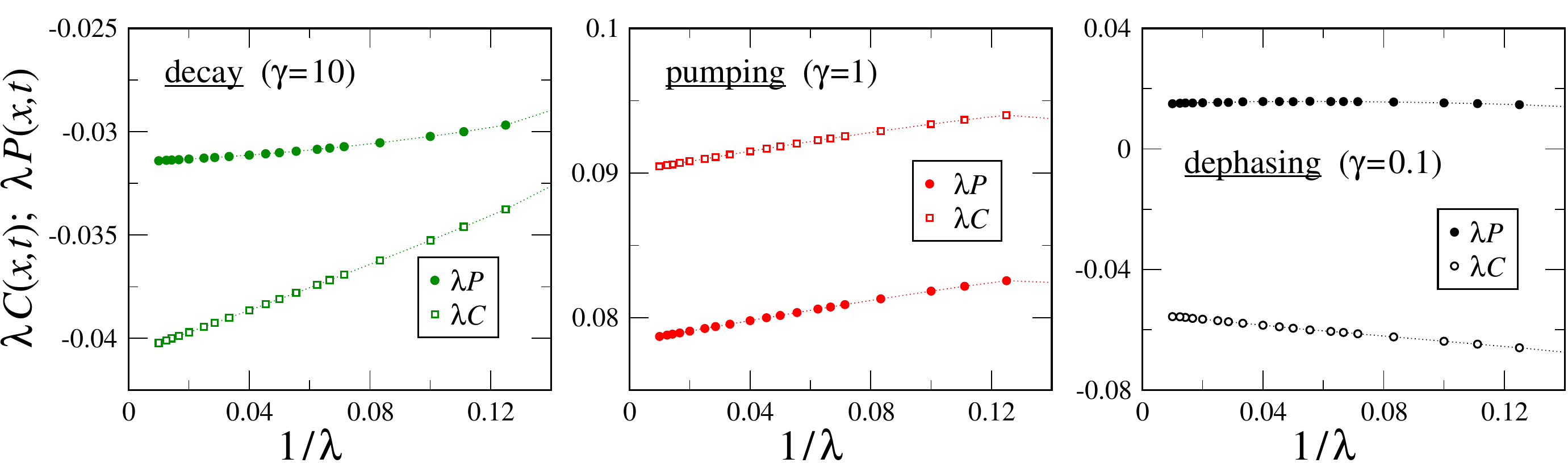}
  \caption{Rescaled correlations $\lambda \, P(x,t)$ (upper panels)
    and $\lambda \, C(x,t)$ (central panels), at fixed rescaled distance
    $x/\lambda = 1$ and initial rescaled time $\tau_i = -10$, for
    the dissipative Kitaev quantum wire in the thermodynamic limit,
    as a function of the scaling variable $\tau$.  
    Analogous scaling behaviors are observed for other values of $x/\lambda$ 
    and $\tau_i$, with different asymptotic scaling functions of $\tau$.    
    Each panel in one of the three columns refers to a specific type
    of dissipation mechanism [see Eq.~\eqref{loppe}]: decay (left),
    pumping (central), and dephasing (right).  The color code stands
    for three rescaled dissipative couplings: $\gamma = 0.1$ (black),
    $\gamma = 1$ (red), and $\gamma=10$ (green).
    Different line styles are for various
    values of $\lambda$, from $8$ to $16$ (see legend).  The lower
    panels show rescaled correlations as a function of $1/\lambda$, up
    to $\lambda = 10^2$, for $\tau=6$ (arrows in the panels above) and
    a given value of $\gamma$ for each panel (see figure).}
\label{fig:PC_TL_diss_tau}
\end{figure*}

Before discussing the effects of dissipation, it is instructive to
present the outcomes of a typical KZ protocol for the unitary dynamics
of the Kitaev quantum wire, without dissipation ($u=0$).  
Figure~\ref{fig:P_TL_nodiss_tau}
shows the time behavior of the fixed-time correlation $P(x,t)$ 
at fixed $x/\lambda$ [see
  Eq.~\eqref{ptf}] during a KZ protocol starting from a fixed
rescaled initial time $\tau_i <0$ (in the figure,
$\tau_i=-10$) and running up to positive values of the rescaled time.
In this way, the Hamiltonian parameter $\bar \mu \equiv \mu-\mu_c$ of
$\hat H_K$ is slowly changed in time starting from an initial value
$\bar \mu_i = \tau_i / \lambda<0$, whose absolute value decreases with
the KZ length scale $\lambda$, through the critical point $\bar\mu=0$
at $t=\tau=0$.  
The various parameters have been rescaled according to Eq.~\eqref{kzG2}:
we set $\varphi = 1$, $\kappa = 1/2$, and plot $\lambda \, P(x,t)$
as a function of $\tau$, for increasing values of $\lambda$,
while keeping the scaling variables $x/\lambda$ and $\tau_i$ constant.  
In Fig.~\ref{fig:P_TL_nodiss_tau} we show results for the rescaled
distance $x/\lambda=1$ only; other values of $x/\lambda$ present analogous
behaviors. Already for values of $\lambda \sim 10$,
the curves approach a non trivial scaling behavior, in
accordance with the general KZ scaling theory for closed systems (see
also the zoom in the upper right inset).  The oscillating behavior for
$\tau > 0$ is likely due to adiabaticity losses, which are ascribable
to the gapless point at $\tau=0$.  The approach to the asymptotic
behavior in the limit $\lambda \to \infty$ is analyzed in the lower
inset, for fixed $\tau = 4.8$, where we collected data up to
$\lambda = 40$. As expected, corrections are suppressed with
a power-law behavior that is compatible with $O(1/\lambda)$.

As stated in the previous sections, the dynamic KZ scaling is also
expected to be independent of the actual value of $\bar \mu_i$, if
this is kept fixed in the dynamic KZ limit.  A numerical verification
of this conjecture is presented in Fig.~\ref{fig:PCG_TL_nodiss_mu},
for a situation similar to that in Fig.~\ref{fig:P_TL_nodiss_tau} but
fixing $\bar \mu_i <0$, rather than $\tau_i < 0$.  Specifically, we
have analyzed the three correlation functions $P(x,t)$, $C(x,t)$, and
$G(x,t)$ [see Eqs.~\eqref{ptf}-\eqref{gntf}] along a KZ protocol where
we fixed the initial condition $\bar \mu_i$.  Even in this case we can
see that, after properly rescaling the various parameters and
observables, the curves nicely approach a scaling behavior, which
appears to be independent of the choice of $\bar \mu_i$.  As
previously discussed, the critical point located at $\bar \mu =
0$ prevents the system from remaining in the instantaneous ground
state.

\begin{figure*}[!t]
  \includegraphics[width=0.95\textwidth]{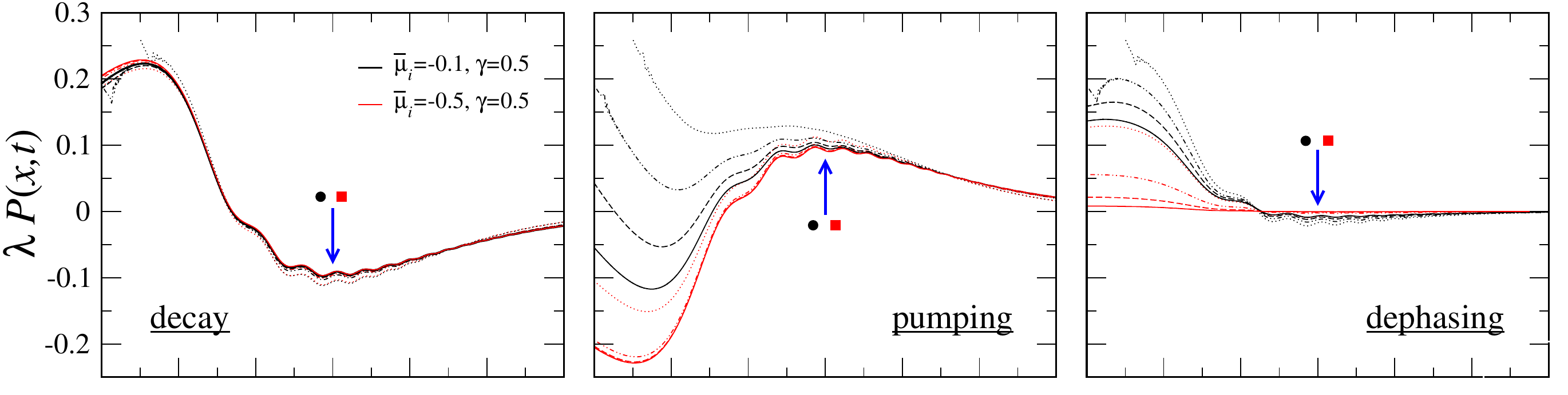}
  \includegraphics[width=0.95\textwidth]{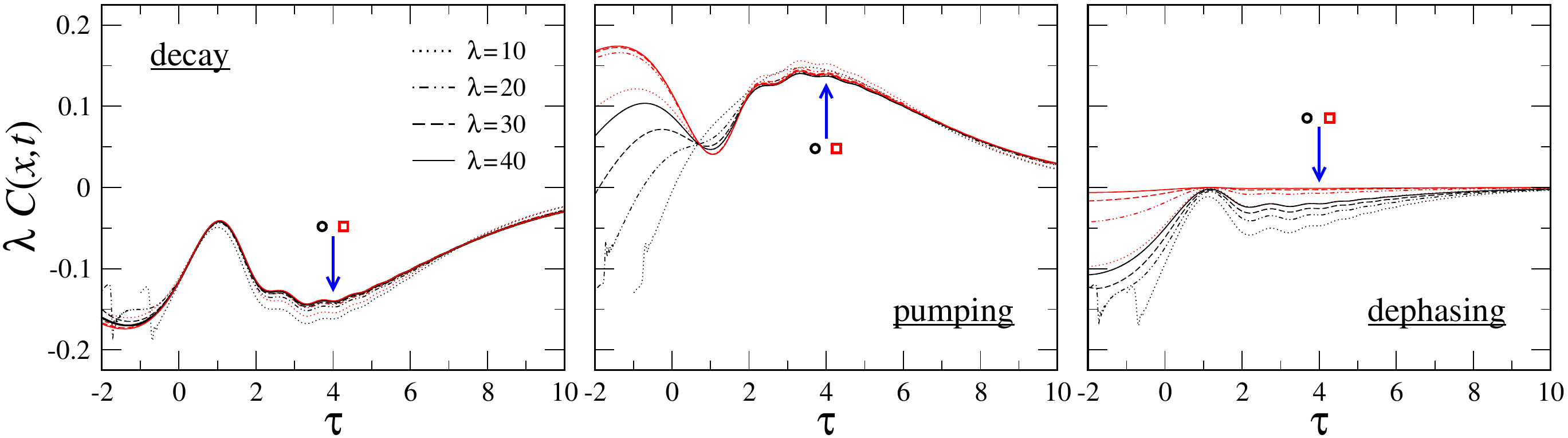}
  \includegraphics[width=0.95\textwidth]{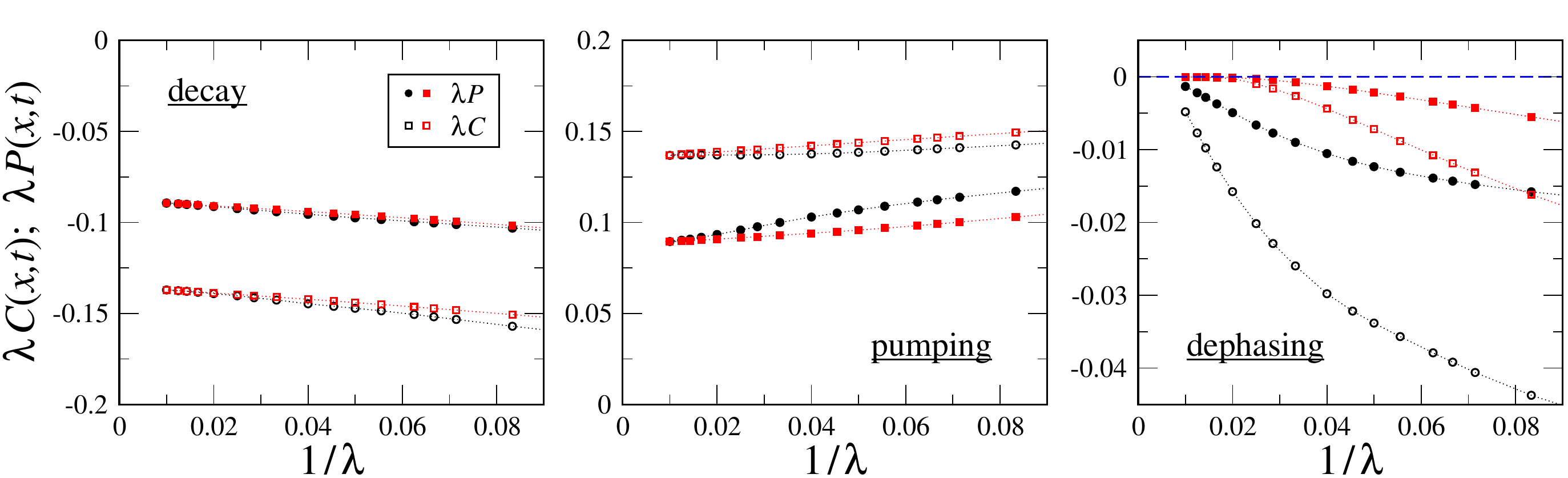}
  \caption{Same kind of analysis as in Fig.~\ref{fig:PC_TL_diss_tau},
    but fixing the initial Hamiltonian parameter $\bar \mu_i$, rather
    than the initial time $\tau_i$. We again show results for $x/\lambda=1$.
    In all the panels we kept the
    rescaled dissipation strength fixed and equal to $\gamma = 0.5$.
    The color code refers to $\bar \mu_i = -0.1$ (black) and to $\bar
    \mu_i= -0.5$ (red), while different line styles stand for various
    values of the length scale $\lambda$, from $10$ to $40$.  The
    lower panels display rescaled correlations $\lambda \, P(x,t)$
    (filled symbols) and $\lambda C(x,t)$ (empty symbols) as a
    function of $1/\lambda$, for $\bar \mu_i = -0.1$ (circles) or
    $\bar \mu_i = -0.5$ (squares), at fixed $\tau=4$ (arrows in the
    above panels).  Panels in the three columns refer to incoherent
    decay (left), pumping (central), and dephasing (right).}
  \label{fig:PC_TL_diss_mu}
\end{figure*}

A more accurate analysis of the independence of the dynamic scaling
functions from the initial condition $\bar \mu_i$ is provided in the
three insets (each for a different correlation function), where we
spotlight the convergence of the rescaled observables with
$\lambda \to \infty$, for a fixed value of $\tau$.  In all the three
cases we observe that the extrapolated asymptotic value seems
to be independent of the two specific $\bar \mu_i$ analyzed.
Notice however that, while for the red data sets ($\bar \mu_i=-0.5$)
the convergence to the asymptotic behavior appears regular and
compatible with a power law $\sim \lambda^{-1}$, the black data
sets ($\bar \mu_i = -0.1$) [especially for $C(x,t)$ and $G(x,t)$]
exhibit oscillations in $1/\lambda$ (at least up to $\lambda=40$),
which should be ascribed to the proximity of the initial
ground state at $\bar\mu_i$ with that at the critical point $\bar
\mu=0$.  We have verified that the above observations hold also for
other values of $\tau$ and for different initial conditions $\bar
\mu_i$ (not shown), with a faster convergence for larger values of
$|\bar \mu_i|$.

\subsubsection{Dissipative KZ dynamics}
\label{sec:TLdissip}

We now turn to a situation where the Hamiltonian Kitaev chain $\hat
H_K$ is coupled to a Markovian bath in the form of either incoherent
particle losses, pumping, or dephasing [the three different types of
  Lindblad operators are reported in Eq.~\eqref{loppe} and are
  supposed to act uniformly over all the sites of the chain].
According to the dynamic KZ scaling framework discussed in
Sec.~\ref{dissint}, an additional scaling variable $\gamma$ associated
with the dissipation strength $u$ needs to be considered, see
Eq.~\eqref{hatgdef}.  In passing we note that the dissipation
parameter $u=\gamma / \lambda$ entering the master
equation~\eqref{lindblaseq} is inversely proportional to $\lambda$,
therefore it needs to be progressively decreased down to zero when
increasing the KZ length scale $\lambda$.  For the two-point
correlations analyzed here, one thus expects the emerging scaling
behavior~\eqref{inflimdisG}.

We proceed as in the previous subsection~\ref{sec:TLunitay} in the
absence of dissipation, and first address KZ protocols where the
Hamiltonian parameter $\bar \mu$ is slowly increased and driven across
a critical point ($\bar \mu=0$), starting from an initial value $\bar
\mu_i<0$ such that the corresponding value of the rescaled time
$\tau_i$ is kept fixed. Results for the correlations $P(x,t)$
and $C(x,t)$, in the presence of either decay, pumping, or dephasing,
are shown in Fig.~\ref{fig:PC_TL_diss_tau}
[analogous outcomes have been obtained for $G(x,t)$---not shown].  
Again, we present results for the particular rescaled distance $x/\lambda=1$ 
and rescaled initial time $\tau_i=-10$;
analogous scaling behaviors are observed for other values 
of $x/\lambda$ and $\tau_i$, but, of course, with different
asymptotic scaling functions of $\tau$.    
  
The upper and central panels
evidence that, after a proper rescaling of the various variables, and
in particular fixing the rescaled dissipation rate $\gamma$ as in
Eq.~\eqref{hatgdef}, the two observables nicely approach a scaling
function with increasing $\lambda$. Of course, the latter function
depends both on the type of dissipation and on $\gamma$. In
particular, numerical data show that the dephasing mechanism appears
to be more effective in destroying this type of correlations: with
increasing $\gamma$, the various curves rapidly decay to a very small
asymptotic value for $\lambda$ large (i.e., green curves for
$\gamma=10$ in the right panels are hardly distinguishable from zero).
On the other hand, for the incoherent decay or pumping, definitely
larger values of $\gamma$ are required to suppress correlations.

The convergence to the asymptotic behavior is analyzed more in depth
in the lower panels for a fixed $\tau$, where we explicitly show the
dependence of the correlation functions on $1/\lambda$, up to $\lambda =100$.
Our data hint at the presence of $1/ \lambda$ power-law corrections,
similarly to what has been observed for the unitary case (compare with
Fig.~\ref{fig:P_TL_nodiss_tau}).

\begin{figure}[!t]
  \includegraphics[width=0.95\columnwidth]{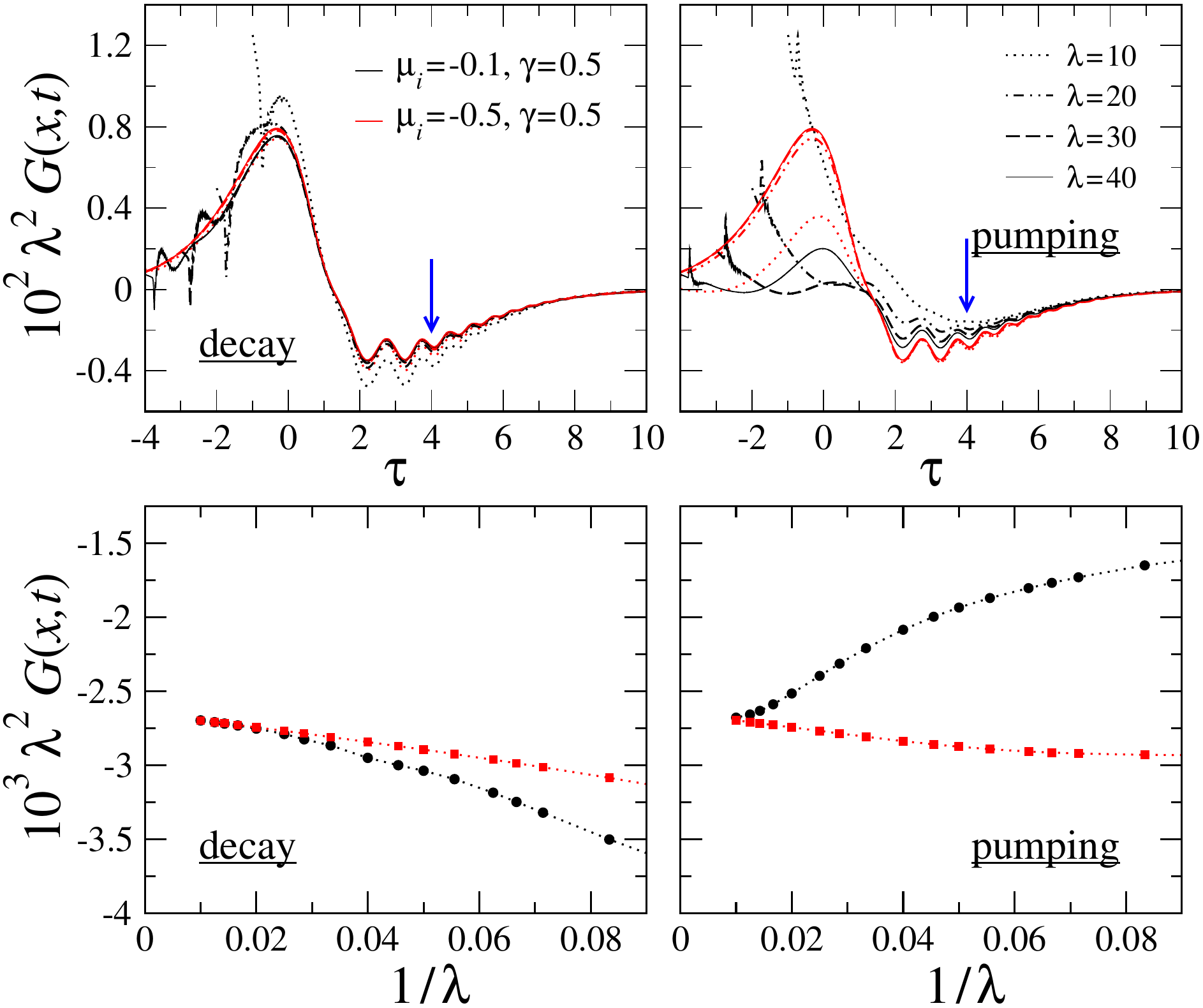}
  \caption{Same analysis as in Fig.~\ref{fig:PC_TL_diss_mu}, but for
    the rescaled correlation $\lambda^2 G(x,t)$.  We have computed it
    the presence of either incoherent decay or pumping, since with
    dephasing we were only able to compute two-point observables
    (while density-density correlations are a four-point
    observable)~\cite{NRV-19-dis}.}
\label{fig:G_TL_diss_mu}
\end{figure}

Even in the presence of dissipation, the dynamic KZ scaling should not
depend on the choice of the initial $\bar \mu_i$, if this is kept
fixed, and thus one expects the scaling behavior reported in
Eqs.~\eqref{sinflimdis}-\eqref{sinflimdisG}.  This has been verified
numerically by fixing $\bar \mu_i<0$, as shown in
Fig.~\ref{fig:PC_TL_diss_mu} for $P(x,t)$ and $C(x,t)$, and in
Fig.~\ref{fig:G_TL_diss_mu} for $G(x,t)$.  The displayed data are for
a specific value of $\gamma=0.5$ and for two different values of
$\bar \mu_i = -0.1$ (black) and $-0.5$ (red).  The various curves in the
upper and central panels stand for different values of $\lambda$. In
all cases we observe that they approach the same asymptotic behavior,
irrespective of the choice of $\bar \mu_i$; the approach becomes
faster, when further increasing $|\bar \mu_i|$ (not shown).  Note
that, for the incoherent decay, this appears to be much faster than
for the other kinds of dissipation (especially for KZ protocols with
$\bar \mu_i= -0.1$, starting close to the critical point $\bar
\mu=0$).  In contrast, as hinted when commenting the upper and middle
right panels of Fig.~\ref{fig:PC_TL_diss_tau} for fixed $\tau_i$ and
different $\gamma$, dephasing seems to be the most disruptive
dissipation mechanism: the red curves in the upper and middle right
panels of Fig.~\ref{fig:PC_TL_diss_mu} (where $|\tau_i|$ is much
larger than for the black ones) are rapidly suppressed with $\lambda$.

The bottom panels of Figs.~\ref{fig:PC_TL_diss_mu}
and~\ref{fig:G_TL_diss_mu}, show the convergence to the asymptotic
behavior with $\lambda$ and for a given $\tau$, which is again
expected to be power law.  Note that the speed with $\lambda$ at which
data for the two $\bar \mu_i$ converge to the same value depends on
the type of dissipation and observable.  In general, we observe that
for decay the convergence is much faster than in the other
cases. Moreover, in the limit $\lambda \to \infty$, while with either
decay or pumping the correlators go toward a non-zero value, with
dephasing the scaling functions (for $\tau$ sufficiently larger than
zero) are compatible with zero.

\begin{figure}[!t]
  \includegraphics[width=0.95\columnwidth]{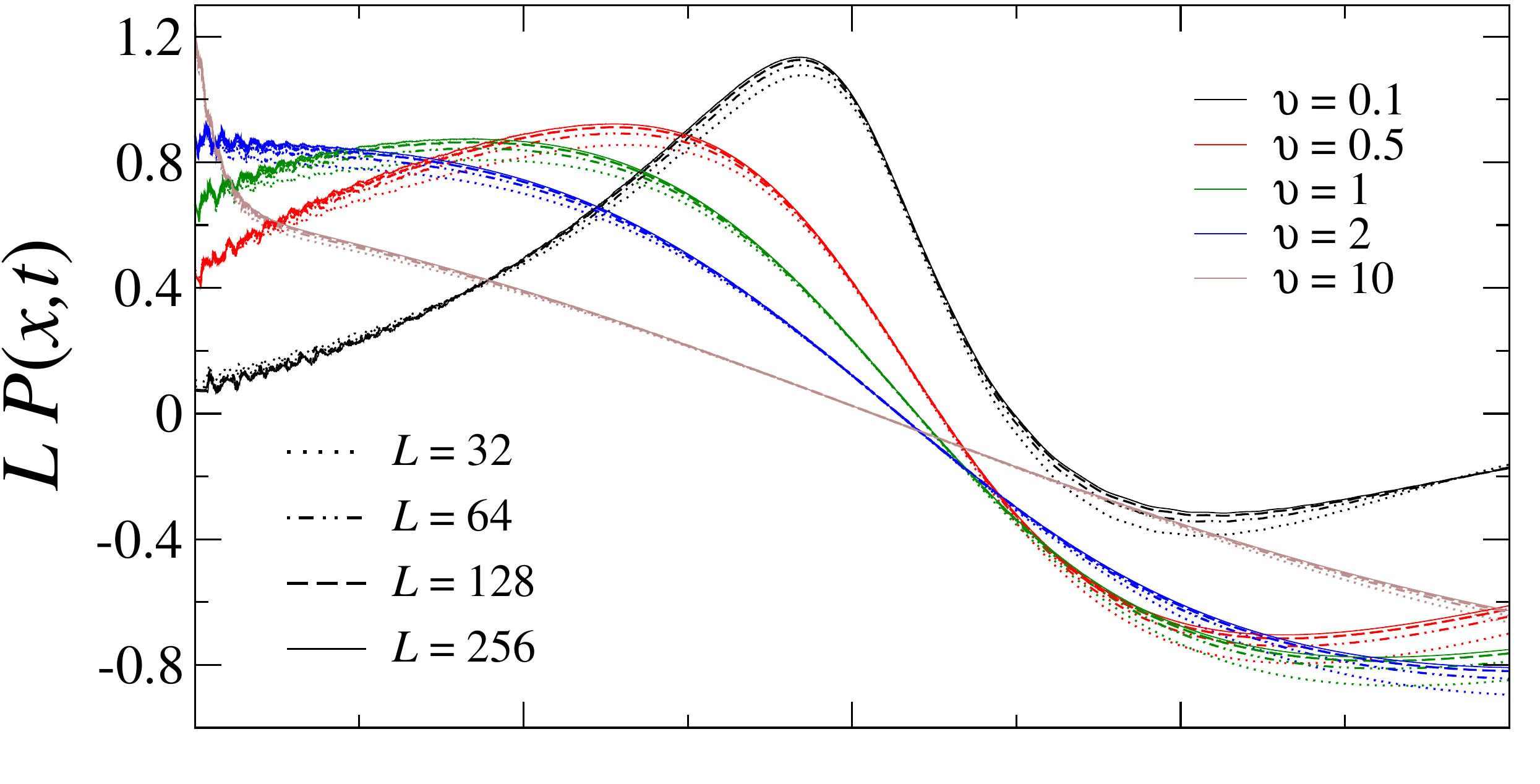}
  \includegraphics[width=0.95\columnwidth]{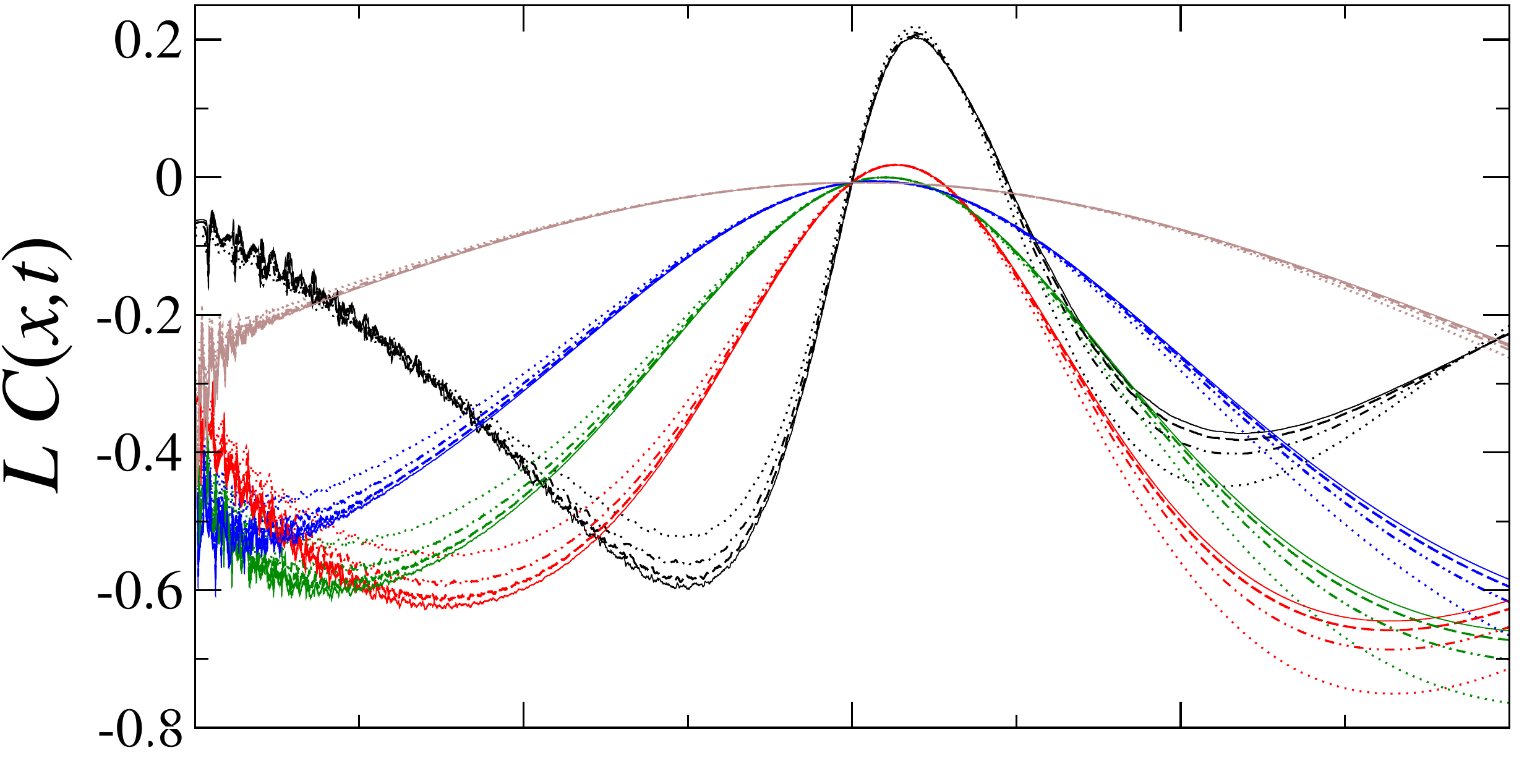}
  \includegraphics[width=0.95\columnwidth]{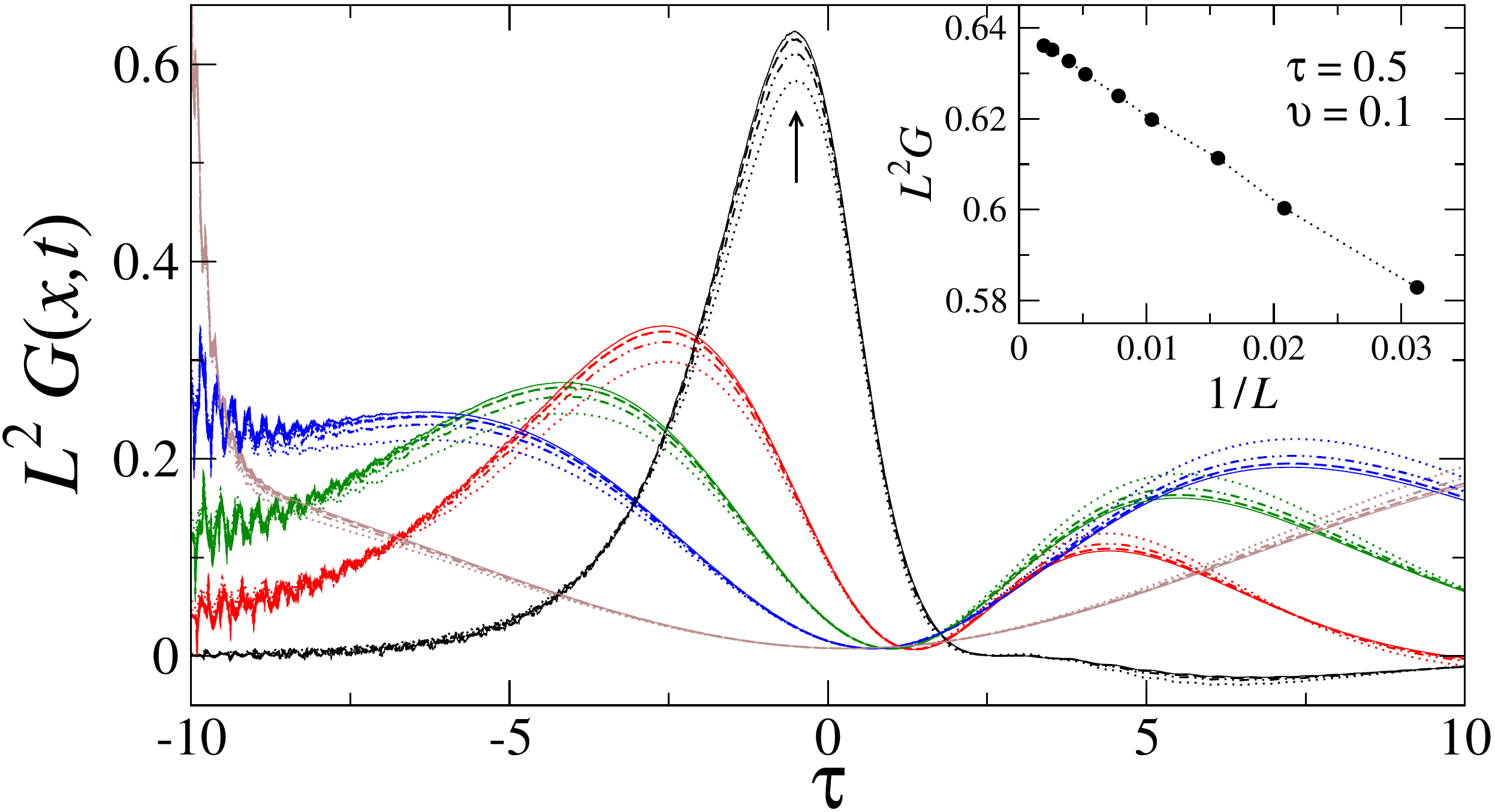}
  \caption{Rescaled correlations $L \, P(x,t)$ (upper panel), $L \,
    C(x,t)$ (central panel), and $L^2 \, G(x,t)$ (lower panel), fixing
    $x/L = 1/4$ (results for other values of $x/L$ show analogous
    behaviors), for the dissipative Kitaev quantum wire with a finite
    length $L$, as a function of the scaling variable $\tau$.  The
    color code corresponds to several values of the inverse KZ speed
    $\upsilon$, while different line styles stand for various system
    sizes $L$ (see legends).  Here we fix the scaling variables
    associated to the initial time ($\tau_i = -10$) and to the
    dissipation ($\gamma_L = 1$), which has been chosen in the form of
    incoherent particle losses.  The inset in the lower panel displays
    the rescaled correlation $L^2 \, G(x,t)$ as a function of $1/L$
    (data up to $L=512$), for fixed $\upsilon = 0.1$ and $\tau =-0.5$
    (arrow in the main panel).}
\label{fig:PCG_scal}
\end{figure}

\subsection{Dynamic KZ finite-size scaling}

We now switch to systems with finite size, and utilize the FSS
framework of Sec.~\ref{fssdis} to analyze the behavior of the
dissipative Kitaev wire undergoing a KZ protocol which crosses the
quantum transition point.  Results for the three fixed-time
correlation functions $P(x,t)$, $C(x,t)$, and $G(x,t)$ are reported in
Fig.~\ref{fig:PCG_scal}, where we analyze their temporal behavior
along a KZ protocol associated with a slow variation of the
Hamiltonian parameter $\bar \mu$ from negative to positive values, in
the presence of incoherent particle losses.  Following the FSS scaling
behavior of Eq.~\eqref{kzfssG2dis}, we kept fixed the ratio $x/L$, the
parameter $\upsilon$ inversely proportional to the speed of the
driving [cf.~Eq.~\eqref{iupsvar}], the rescaled dissipation strength
$\gamma_L$ [cf.~Eq.~\eqref{gammadef}], and the initial rescaled time
$\tau_i <0$.  Note that, in the FSS framework, the dissipation
strength $u=\gamma_L / L$ entering the master
equation~\eqref{lindblaseq} is inversely proportional to the system
size, thus scaling down to zero in the limit $L\to \infty$.

\begin{figure}[!t]
  \includegraphics[width=0.98\columnwidth]{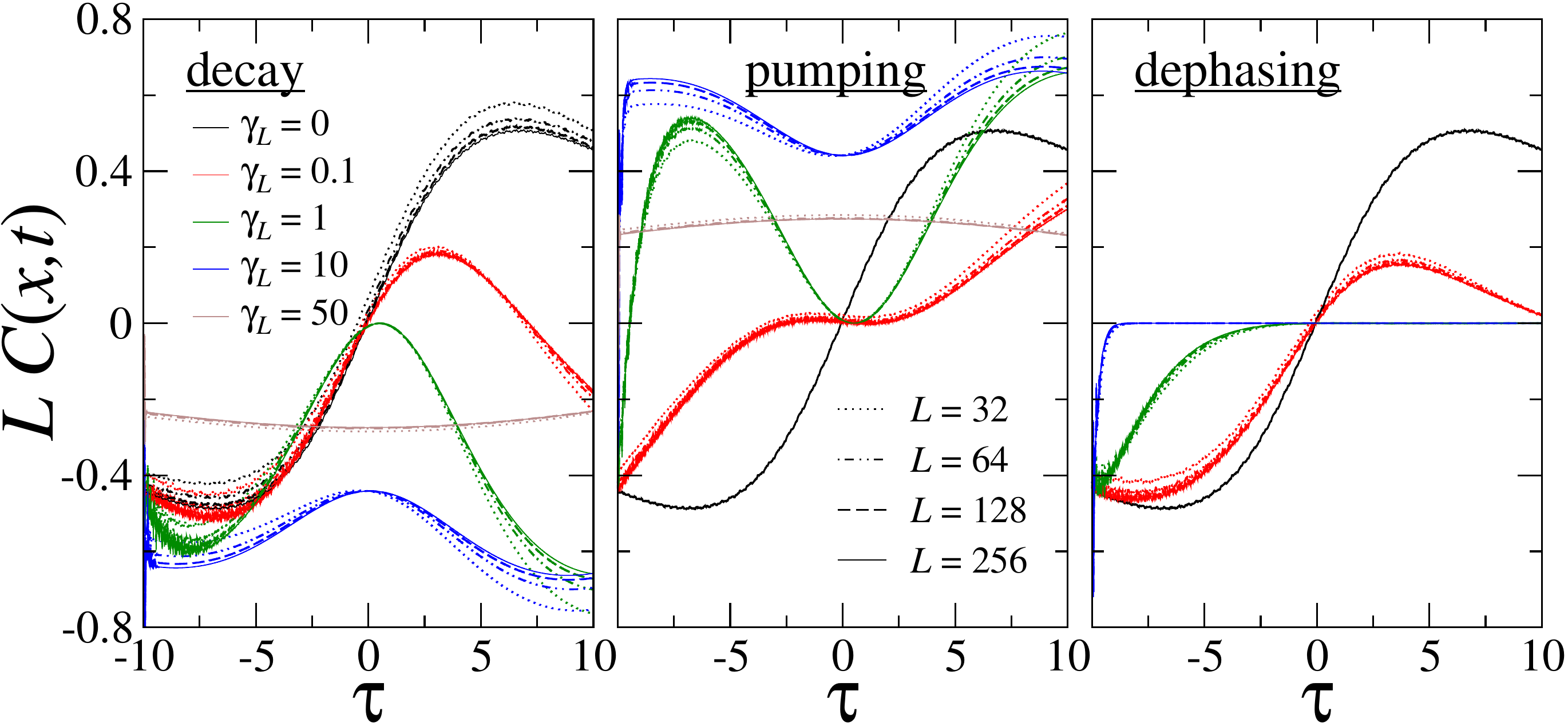} 
  \caption{Same as in Fig.~\ref{fig:PCG_scal}, but for the correlation
    $C(x,t)$, for dissipation given by decay (left), pumping (middle),
    or dephasing (right).  The color code corresponds to several
    values of $\gamma_L$, while different line styles stand for
    various values of $L$ (see legends).  Black curves are for
    $\gamma_L=0$ (in the middle and right panels we replot the same
    curve corresponding to $L=256$, for reference).  Here we fix
    $\tau_i = -10$ and $\upsilon = 1$.}
\label{fig:PC_scal_deph}
\end{figure}

In all cases, the rescaled correlations nicely approach a scaling
function with increasing $L$, as predicted by the scaling law in
Eq.~\eqref{kzfssG2dis}.  The finite-size approach to the asymptotic
behavior is compatible with a $L^{-1}$ behavior, as highlighted in the
inset of the lower panel of Fig.~\ref{fig:PCG_scal} at fixed $\tau$.
We also observe that, for small values of $\tau$, the scaling curves
develop complex nonanalytic spikes in $\tau$, whose magnitude and
frequency increase with $L$, similarly to other dynamic situations as
after sudden quenches~\cite{NRV-19-dis, RV-19-dis}; for larger $\tau$
dissipation tends to smear those apparent singularities.

To shed light on the effects of the system-bath coupling, in
Fig.~\ref{fig:PC_scal_deph} we have analyzed the correlation $C(x,t)$
for the three different types of dissipation of Eq.~\eqref{loppe}, and
for varying rescaled strength $\gamma_L$ as indicated in the legends.
We have also reported the KZ behavior in the unitary case (black
curves), to be compared with that in the presence of an environmental
interaction (colored curves).  Besides the nice convergence to a
scaling function for $L \to \infty$, we observe that dephasing
appears to be more effective in destroying correlations, since, with
increasing $\gamma_L$, the curves rapidly approach the zero value in
time; in contrast, for both incoherent decay and incoherent pumping,
definitely larger values of $\gamma_L$ are required to suppress
correlations.  Moreover, in the presence of pumping, even a tiny
amount of dissipation is capable to drive the system far from the
equilibrium state in the $\tau<0$ side (see also the discussion in
Sec.~\ref{sec:TLdissip}).

\begin{figure}[!t]
  \includegraphics[width=\columnwidth]{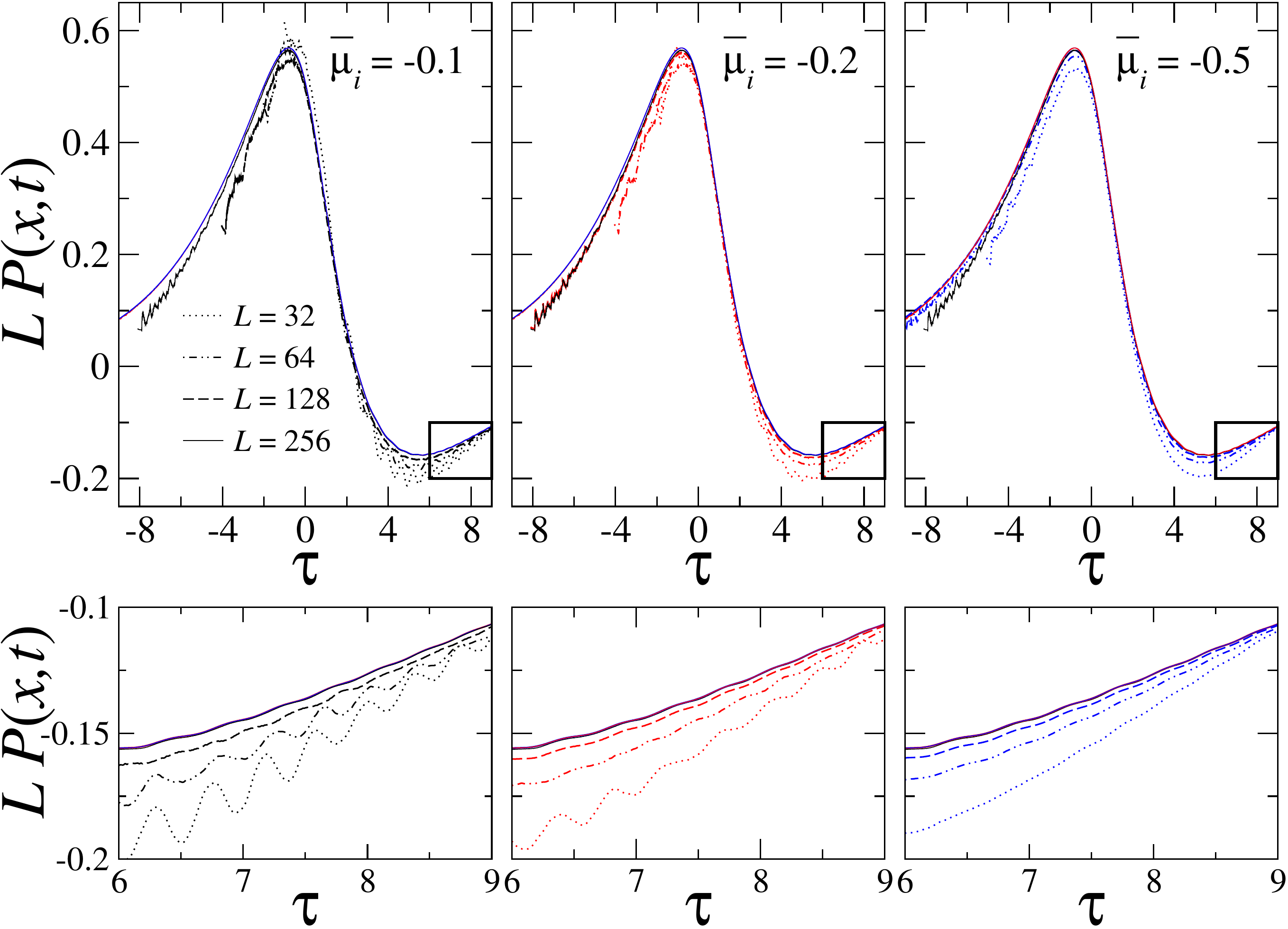} 
  \caption{The rescaled correlation $L \, P(x,t)$ with $x/L = 1/4$ as
    a function of $\tau$ (results for other values of $x/L$ show
    analogous behaviors), fixing the initial Hamiltonian parameter
    $\bar \mu_i = -0.1$ (left panels, black curves), $-0.2$ (central
    panels, red curves), and $-0.5$ (right panels, blue curves).
    Different line styles stand for various $L$ (see legend).  The
    lower panels show magnifications of the upper ones, for
    $6 \leq \tau \leq 9$.  Here we fix $\upsilon = 0.1$, and
    $\gamma_L = 1$ with dissipation given by incoherent decay.
    Note that the three continuous curves, corresponding to the
    largest size $L=256$ and different $\bar \mu_i$, are plotted
    in each of the six panels and cannot be distinguished
    on the scale of the plots reported.}
\label{P_MuIn}
\end{figure}

We have also numerically verified that, analogously to the dynamic
scaling behavior in the infinite-volume limit, the dynamic FSS
functions do not depend on the initial Hamiltonian parameter
$\bar{\mu}_i<0$ if this is kept fixed in the dynamic KZ limit [see
  Eq.~\eqref{kzfssG3}].  Figure~\ref{P_MuIn} reports the
behavior of the rescaled correlation $L \, P(x,t)$ as a function of
$\tau$, for three different choices of $\bar{\mu}_i$, and for
dissipation provided by incoherent decay at fixed $\gamma_L$.  The
various upper panels show convergence with $L$ to a scaling function,
which appears to be the same: bottom panels unveil how tiny
discrepancies in the temporal behavior, starting from different
$\bar \mu_i$, can be suppressed in the large-$L$ limit.

\section{Summary and conclusions}
\label{conclu}

We have investigated the interplay between coherent and dissipative
drivings in the dynamics of quantum may-body systems subject to KZ
protocols across continuous quantum transitions, starting from the
gapped disordered phase, that is when one Hamiltonian parameter is
slowly driven across its critical value, for example with a linear
dependence on time.  Specifically, the issue we have addressed here is
whether, and under which conditions, open dissipative systems can
develop a universal dynamic scaling regime similar to that shown by
closed systems at quantum transitions, characterized by asymptotic
homogeneous scaling laws.  To this purpose we have focused on a class of
dissipative mechanisms, whose dynamics can be reliably described
through a Lindblad master equation governing the time evolution of the
density matrix of the open system.

The perturbation arising from the dissipation turns out to be relevant
at the quantum transition~\cite{NRV-19-dis, RV-19-dis}.  This implies
that open systems cannot develop asymptotic dynamic scaling behaviors
controlled by the universality class of the quantum transition when
keeping the dissipation decay rate $u$ finite and fixed in the
critical limit of the Hamiltonian parameter.  This is analogous to the
effect of any relevant RG perturbation close to a critical
point~\cite{Sachdev-book}, like the temperature, which makes the
system run away from criticality.  Nevertheless, we argue that a
dynamic KZ scaling limit exists in the presence of a sufficiently weak
dissipation. Such a scaling limit, controlled by the universality
class of the quantum transition, arises in a regime of weak
dissipation. Indeed, it requires a tuning of the dissipative
interactions, and in particular of the decay-rate parameter $u$ of the
Lindblad master equation describing the evolution of the density
matrix, cf.  Eq.~(\ref{lindblaseq}). The decay rate $u$ must decrease
as $u\sim t_s^{-\kappa}$ when increasing the time scale $t_s$ of the
KZ protocol, where the positive exponent $\kappa = z/(y_\mu+z)<1$
depends on the dynamic exponent $z$ and the RG dimension $y_\mu$ of
the driving Hamiltonian parameter (usually related to the
correlation-length exponent exponent $\nu$ by $\nu=y_\mu^{-1}$).  The
resulting dynamic KZ scaling laws, allowing for the presence of
dissipation, provide a unique framework to discuss te interplay
between (critical) coherent and dissipative drivings.

The dynamic KZ scaling scenario has been checked within fermionic
wires, cf.~Eq.~\eqref{kitaev2}, in the presence of homogeneous
dissipation due to local incoherent pumping, decay and dephasing,
which are described by the Lindblad operators reported in
Eq.~\eqref{loppe}. The particularly convenient choice of this model
enables to scale its complexity linearly with its size, allowing
to simulate the exact dissipative dynamics of systems with thousands
of sites; we have thus elected it as a testbed for accurate
numerical investigations of the many-body Lindblad master equation.
Our numerical analysis ultimately supports the
phenomenological dynamic KZ scaling framework addressing the
competition between coherent dynamics and dissipation at a continuous
quantum transition.

We believe that, in the near future, it will be also possible to
address and verify this scenario through suitably engineered experiments
with ultracold atoms or cavity-QED technology aimed at realizing 
and controlling driven-dissipative quantum many-body systems
(see, e.g., Ref~\cite{TNDTT-17}).

It would be tempting to investigate and carefully verify our dynamic
KZ scaling in other quantum dissipative systems, such as Ising-like
quantum spin models.  To that purpose, given the difficulties in
finding a numerical solution to the Lindblad master equation for a
generic many-body problem~\eqref{lindblaseq}, a FSS framework should
be adopted as the primary setting, due to the relatively small system
sizes that could be reached and the substantial impossibility to
address infinite-volume systems (this would be the case, e.g., for
the standard quantum Ising chain, with realistic local dissipation
related to the spin operators). In that respect, an interesting
issue would be to extend the dynamic KZ scaling to protocols across
first-order quantum transitions (e.g.~in the quantum Ising chain in a
transverse plus longitudinal field).  The exponentially closing gap
between the two lowest states of the ordered phase might be relevant
and new features may become apparent already for systems with $O(10)$
spins, such as the sensitivity on the type of boundary 
conditions~\cite{CNPV-14, PRV-18c}.

\appendix

\section{Similarities and differences between the Kitaev 
wire and the quantum Ising chain}
\label{app:KitaevIsing}

In Sec.~\ref{Kitaevmo} we stated that the Kitaev quantum wire
described by the Hamiltonian~\eqref{kitaev2} undergoes a continuous
quantum transition in the same universality class of the quantum Ising
chain.  The similarities between the two models can be put on a formal
ground by means of a Jordan-Wigner transformation, which maps the
spinless fermions into spin-$1/2$ operators:
\begin{equation}
  \hat \sigma^{\pm}_j = \exp \bigg( i \pi \sum_{\ell < j} \hat n_\ell
  \bigg) \hat c^{\pm}_j \,.
  \label{eq:jwt}
\end{equation}
Here $\hat \sigma^\pm_j = \tfrac12 \big( \hat \sigma^x_j \pm i \hat
\sigma^y_j \big)$ are the spin-$1/2$ raising/lowering operators and
$\hat \sigma^\alpha_j$ ($\alpha = x,y,z$) denote the usual Pauli
matrices associated to site $j$ in the chain.

Indeed, it can be easily shown that, neglecting boundary terms, the
above transformation maps $\hat H_{\rm K}$ of Eq.~\eqref{kitaev2} into
the XY chain $(\delta \neq 0)$:
\begin{equation}
  \hat H_{\rm XY} = - t \sum_j \bigg[ \frac{1+\delta}{2} \hat
    \sigma^x_j \hat \sigma^x_{j+1}+ \frac{1-\delta}{2} \hat \sigma^y_j
    \hat \sigma^y_{j+1} + \frac{\mu}{2t} \hat \sigma^z_j \bigg] \, .
\end{equation}
In particular, for $t = \delta = 1$, the corresponding
spin model coincides with the quantum Ising chain 
\begin{equation}
  \hat H_{\rm Is} = - \sum_j \big( \hat \sigma^x_j \hat \sigma^x_{j+1}
  + g \hat \sigma^z_j \big) \,,
\end{equation}
with $g=-\mu/2$.

It is however crucial to stress that the boundary conditions play an
important role in this mapping.  As a matter of fact, the non-local
Jordan-Wigner transformation of the Ising chain with periodic or
antiperiodic boundary conditions does not map into the fermionic
model~\eqref{kitaev2} with periodic or antiperiodic boundary
conditions.  Indeed further considerations apply~\cite{Katsura-62,
  Pfeuty-70}, leading to a less straightforward correspondence, which
also depends on the parity of the particle number eigenvalue (see
below).  Therefore, although the bulk behaviors of the above models in
the infinite-volume limit (and thus their phase diagram) are
analogous, the resulting FSS functions are different, since they
subtly depend on the choice of the boundary conditions.

Even more, the Kitaev quantum wire with antiperiodic boundary
conditions, explicitly studied in this paper, turns out to be gapped
in both of the phases separated by the quantum transition at
$\mu_c=-2$.  Indeed, the energy difference $\Delta$ of the two lowest
states is given by
\begin{equation}
  \Delta = \sqrt{\bar\mu^2 + 4 (2-\bar\mu) \, [1-{\rm cos}(\pi/L)]}\,,
\end{equation}
where $\bar\mu=\mu-\mu_c$, such that
\begin{equation}
  \Delta = \left\{ \begin{array}{ll} \displaystyle
    |\bar\mu| + {\pi^2 (2-\bar\mu)\over |\bar\mu| L^2} + O(L^{-4}) 
    & {\rm for} \quad |\bar\mu| > 0 \,, \vspace*{1mm} \\
    \displaystyle {2\pi \over L} + O(L^{-3}) & {\rm for} \quad |\bar\mu| = 0 \,.
  \end{array} \right.
\end{equation}
Therefore, the Kitaev quantum wire studied here does not exhibit the
lowest-state degeneracy of the ordered phase of the quantum Ising
chain (i.e., the exponential suppression of the gap with increasing
$L$).  The reason for that substantial discrepancy resides in the fact
that the Hilbert space of the Kitaev quantum wire with antiperiodic
boundary conditions alone is restricted with respect to that of the
quantum Ising chain, so that it is not possible to restore the
competition between the two vacua belonging to the symmetric/antisymmetric
sectors of the Ising model~\cite{Katsura-62, Kitaev-01, CPV-14}.

The ultimate reason why we prefer to stick with the Kitaev quantum
wire is twofold: {\it i)} the dissipation that we consider in this
paper is more naturally defined for Fermi lattice gases; {\it ii)} the
dissipative fermionic decay/pumping mechanisms cannot be mapped into
simple spin operators, due to the presence of a nonlocal string
operator in the transformation~\eqref{eq:jwt}.
In this respect, simulating a conventional quantum Ising chain
with local dissipation in the form of spin losses
($\hat L_{{\rm l},j} = \hat \sigma^-_j$) or pumping
($\hat L_{{\rm p},j} = \hat \sigma^+_j$) would prevent one from exploiting
the particularly simple solvability of the Kitaev model with a
polynomial amount of resources, due to the appearance of Jordan-Wigner
strings when mapping the term $\hat L_j \rho \hat L^\dagger_j$
of the Lindblad master equation in fermionic language~\cite{KMSFR-17}.

In light of this, it is finally worth mentioning that, although we
have only shown numerical results for KZ protocols where the
Hamiltonian parameter $\mu$ is linearly driven in time from an initial
value $\mu_i < \mu_c$ to a final value $\mu_f > \mu_c$, there is no
reason to expect qualitative differences when reverting the protocol,
i.e., starting from $\mu_i > \mu_c$ and ending into $\mu_f < \mu_c$.
The reason resides in the fact that the Kitaev chain with antiperiodic
boundary conditions is gapped in both
phases on the left and on the right of the quantum transition point
$\mu_c$, and thus the evolution arising from slow changes of $\mu$ is
essentially adiabatic, from any $\bar\mu_i$ far from criticality to
the relevant scaling interval around $\mu_c$.  The situation may
change for the above mentioned quantum Ising chain, since one
of the two phases is ordered and presents a double degeneracy.

\end{document}